\documentclass[11pt, a4paper]{article}

\pdfoutput=1
\usepackage{jheppub}
\usepackage{etoolbox}
\usepackage{amsmath,amsfonts,amssymb,bbm, mathtools, MnSymbol}
\usepackage{hyperref, csquotes} 
\usepackage{graphicx, color}
\usepackage{array}
\usepackage{tikz,pgf}
\usetikzlibrary{calc,angles}
\usepackage{cancel}
\usepackage{caption}
\usepackage{subcaption}
\usepackage{multirow}
\usepackage{tablefootnote}
\usepackage{comment}
\usetikzlibrary{patterns}
\usepackage{physics,tensor}
\usepackage{dirtytalk}
\usepackage{orcidlink}

\renewcommand{\eqref}[1]{Eq.\,(\ref{#1})}

\newcommand{\R}{\mathrm{R}}
\newcommand{\J}{\mathrm{J}}
\renewcommand{\P}{\mathrm{P}}

\definecolor{jred}{rgb}{0.8,0,0}
\definecolor{jgreen}{rgb}{0,0.7,0}
\definecolor{jblue}{rgb}{0,0,0.8}

\newcommand{\jhepauthor}[4][]{%
  \author[#1]{#2\,\orcidlink{#3}}%
  \emailAdd{#4}}

\begin{document}

\title{Quasinormal-mode stability bounds in gravitational EFT: resolvability, predictivity, and positivity}

\jhepauthor[a, b]{Leonardo Juliano}{0009-0009-0731-2297}{leonardo.juliano@sns.it}
\jhepauthor[a, b]{Enrico Trincherini}{0000-0001-5025-0292}{enrico.trincherini@sns.it}
\affiliation[a]{Scuola Normale Superiore, Pisa, Italy}
\affiliation[b]{INFN Sezione di Pisa, Largo Pontecorvo 3, 56127 Pisa, Italy\\}

\begin{abstract}{It has been proposed that consistency limits how slowly black holes can ring down in extensions of General Relativity: effective-field-theory (EFT) corrections may lengthen the decay time of a quasinormal mode by at most about one oscillation period, $\Re[\omega_{\rm GR}]\,\delta\tau\lesssim1$. We argue that, under standard power counting, this condition yields no new, model-independent constraint on gravitational EFTs. First, its motivation does not extend straightforwardly to black holes: the inverse oscillation frequency is not a universal, detector-independent uncertainty in the damping time of a classical waveform. Our main argument, however, rests on eikonal power counting alone. The decay-time shift grows with the angular momentum $\ell$ of the mode, but so does the weight of higher-derivative operators. When curvature and derivatives are controlled by a single scale $\Lambda$, as in every known ultraviolet completion, the EFT around a black hole of radius $r_s$ remains predictive up to $\ell\sim(r_s\Lambda)^2$, and the shift becomes of order one only there, where an infinite tower of operators contributes at the same order.  A resolvable shift within the predictive regime requires either large numerical coefficients or a two-scale power counting, which has no known ultraviolet completion. In both cases the quartic-curvature operators dominate, and the condition coincides parametrically with the positivity bounds that flat-space graviton amplitudes impose under standard assumptions on the ultraviolet completion. Only in scalar--tensor theories could the two-scale window probe beyond positivity, and then at most one family of dimension-ten operators. The axial sector of Einstein--dilaton--Gauss--Bonnet gravity and Einstein--Maxwell theory illustrate these results.}
\end{abstract}

\setcounter{tocdepth}{2}

\maketitle

\section{Introduction}
After two black holes merge, the remnant settles to a stationary geometry by ringing down, emitting a superposition of damped tones, the quasi-normal modes (QNMs), each labeled by a complex frequency $\omega$ whose real part sets the oscillation frequency and whose imaginary part the decay time $\tau$\footnote{For a recent review, see \cite{Berti2025}.}. For a Kerr Black hole in General Relativity (GR) these frequencies are fixed by the mass and spin of the remnant. A measured departure from that spectrum --- once astrophysical environments and waveform systematics are under control --- is therefore a window into gravitational dynamics beyond GR. 

The natural language for describing such departures is Effective Field Theory (EFT). If the graviton is the only light degree of freedom, one supplements the Einstein--Hilbert action with every local, diffeomorphism-invariant operator built from the curvature and its derivatives (GREFT). Two scales control the expansion: $\Lambda$, suppressing each extra power of curvature, and $\Lambda_\nabla$, each extra derivative. Neither of them is, a priori, the mass of the heavy states that have been integrated out, nor the scale at which a given scattering process becomes strongly coupled; keeping the three distinct will matter below.  Additional light degrees of freedom, where present, widen the operator basis without drastically changing our conclusions: we return to the scalar-tensor case in the section \ref{scalar-tensor}. 

The Wilson coefficients of the EFT operators are not free. Analyticity and unitarity of flat-space amplitudes already carve out the allowed region \cite{Adams2006, Cheung_2014, Arkani_Hamed2021, Bellazzini2022, Bellazzini2023, Caron-Huot:2022ugt}. The absence of resolvable time advances for waves propagating on non-trivial backgrounds, black holes in particular, has been used to the same end \cite{Camanho16, Papallo_2015, Benakli_2016, Hollowood_2016, Goon_2017, de_Rham2007, de_Rham2112, Chen2112, Serra2205, chen2024surfinppwavesgoodvibrations, Nie2410}, although in four dimensions it is a subtler requirement than it may seem \cite{Bucciotti:2026svg}.

To these, \cite{Melville2401} has recently added a consistency condition of a different kind, phrased directly in terms of the ringdown. Writing the EFT shift\footnote{For relevant literature in the context of EFT correction to Schwarzschild QNMs, see: \cite{Franciolini1902, Cardoso_2018, Endlich_2017, Silva_2023, Nakashi_2026, Cardoso_2019, McManus_2019, Mukohyama_2022, de_Rham_2020, Mukohyama2204, Bryant_2021}.} in the damping time as $\delta\tau \equiv \tau^{\text{EFT}} - \tau^{\text{GR}}$, the proposal is that
\begin{equation}\label{eq:introbound}
\Re[\omega_{\rm GR}]\, \delta\tau \lesssim 1\,.
\end{equation}
In words: an EFT-corrected Black hole should ring down at least as fast as its GR counterpart, and if it rings down more slowly, by no more than a single oscillation period. Three arguments are offered in support: that integrating out heavy states drives propagator poles deeper into the lower half-plane; that \eqref{eq:introbound} is the ringdown analogue of the time-advance condition; and that, in the cases examined in \cite{Melville2401}, it tracks the positivity and causality bounds and occasionally sharpens them. 

Our main result is that, under standard EFT power-counting assumptions, \eqref{eq:introbound} does not yield a new, model-independent constraint on the Wilson coefficients. When the curvature and derivative expansions are controlled by a single scale, $\Lambda\sim\Lambda_{\nabla}$ --- the only power counting with a known ultraviolet completion --- there is no parametric regime in which $\Re[\omega_{\rm GR}]\,\delta\tau$ is of order one or larger while a generic EFT remains predictive. Here and below, by a generic EFT we mean a symmetry-complete operator basis with coefficients of natural size in the assumed power counting and without tuned cancellations. At angular momenta well below the EFT breakdown scale, the decay-time shift is parametrically smaller than unity; increasing $\ell$ enhances the shift, but it becomes order one only at the boundary where an infinite tower of higher-derivative operators contributes at the same order. The one exception is the two-scale counting $\Lambda_{\nabla}\gg\Lambda$ which, besides being excluded by positivity, does not correspond to any known UV completion. There, the shift does become parametrically resolvable below the cutoff. Despite this, \eqref{eq:introbound} adds almost nothing to existing constraints for GREFT as the resulting bounds overlap parametrically with positivity.

The same analysis carries over beyond pure gravity EFT. The only enhanced case is obtained in the two scale scenario for a scalar tensor theory, where the only probed operator sits at dimension 10. We study Einstein--dilaton--Gauss--Bonnet (EdGB) gravity in the axial sector as an explicit example: a formal double-scaling limit can produce an upward pole displacement large enough to cross the proposed threshold, but the required angular momentum grows parametrically beyond the EFT validity range. We also comment on the application of the same reasoning to Einstein--Maxwell EFTs, where we can compare with the $\ell = 2$ numerical results of \cite{Boyce2025}; the conclusion is unchanged. 

We establish this statement by extending the leading WKB analysis to effective radial equations with frequency- and angular-momentum-dependent corrections. For a Black hole of Schwarzschild radius $r_s$, perturbativity of the linearized equations in GREFT gives the conservative upper estimate
\begin{equation}
    \ell_{\max}
    \sim
    \begin{cases}
        \ (r_s\Lambda)^2
        & \text{for}\quad \Lambda_\nabla \sim \Lambda\,,\\[6pt]
        \ \min\left\{\,(r_s\Lambda)^2\,\dfrac{\Lambda_\nabla}{\Lambda}\,,\;(r_s\Lambda)^3\right\}
        & \text{for}\quad \Lambda_\nabla \gg \Lambda\,.
    \end{cases}
    \label{eq:ell-max-intro}
\end{equation}
Evaluating the dimensionless shift at this upper limit makes the statement quantitative: for $\Lambda_{\nabla}\sim\Lambda$ it reaches at most order one, and only at the limit itself; for $\Lambda_{\nabla}\gg\Lambda$ it can grow as large as $\min\{(\Lambda_{\nabla}/\Lambda)^3,(r_s\Lambda)^3\}$, in that window, which we analyze in Section~\ref{sec:4}, the proposed bound reduces to the positivity of the two dimension-8 coefficients, already required by causality of flat-space amplitudes \cite{Caron-Huot:2022ugt}. Therefore, it carries no information beyond
existing constraints.

We also revisit the motivations offered for \eqref{eq:introbound}. The tolerance of one oscillation period is borrowed from the time-advance condition, where it reflects the width of the probe wave packet. A ringdown, however, is a macroscopic, phase-coherent waveform, and $1/\Re[\omega]$ is not a fundamental uncertainty in its damping time: the precision with which a pole or waveform parameter can be inferred depends on the signal-to-noise ratio, occupation number, number of observed events, and measurement protocol.
These observations weaken the proposed motivations, although our principal conclusion relies only on the EFT scaling argument.

The paper is organised as follows. Section~\ref{sec:1} states the stability bound and examines its motivations. Section~\ref{sec:4} builds the WKB framework and works through GREFT and scalar-tensor theories, the latter including the EdGB case study. Section~\ref{sec:EM} turns to Einstein-Maxwell theory, and Section~\ref{sec:Conclusion} concludes. Three appendices collect the WKB treatment of coupled systems (Appendix~\ref{app:A}), the power-counting behind the validity-scale tables (Appendix~\ref{app: B}), and the Lyapunov-exponent computation used for EdGB (Appendix~\ref{app: C}).

\section{Black hole stability bound}\label{sec:1}
\subsection{Stability conjecture} \label{sec:1.2}
In the context of black holes' physics, quasi-normal modes (QNMs) are damped oscillations that arise from the perturbation of the black holes' background metric. Those oscillations are characterized by a frequency and a decay time $\tau$ and are usually described in terms of a single complex variable $\omega$, whose real part describes the oscillatory behavior, and the imaginary part the damping, $\tau := - 2\pi / \Im[\omega]$. Both frequency and decay time are modified by irrelevant operators added to the Einstein-Hilbert action, in particular:
\begin{equation}
    \tau_{\text{EFT}} = \tau_{\text{GR}} + \delta \tau
\end{equation}
where we refer to $\delta \tau$ as \textit{decay-time shift}. QNMs can also be thought of as poles of the Green's function solving the differential equation
\begin{equation}
    \left( \frac{d^2}{dr_*^2} + \omega^2 - V(r_*) \right) G(\omega, r_*, r_*') = - \delta(r_*- r_*')
\end{equation}
with oscillatory boundary conditions at infinity and the horizon \cite{Leaver:1986gd, Berti2009, Berti2025}. $V(r_*)$ is a potential that depends on the specific theory and the type of perturbation (scalar, vector, etc.). Linear QNMs are classified, using three \say{quantum numbers} $(\ell, m, n)$ respectively the angular momentum, the azimuthal component of angular momentum and the overtone. In what follows we will ignore the azimuthal and the overtone quantum numbers, setting them to zero for simplicity. 

\begin{figure} 
    \centering
    \includegraphics[width=\linewidth]{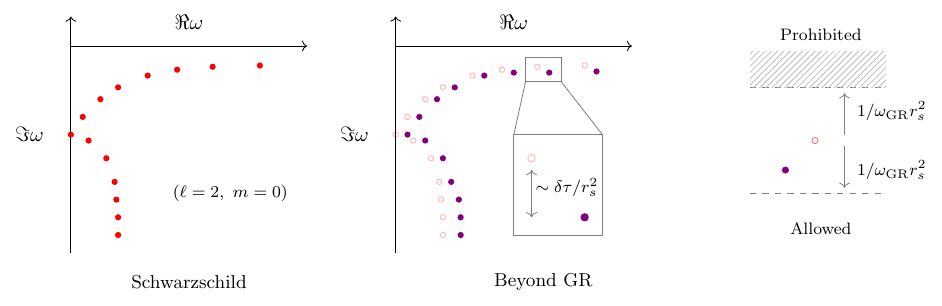}
    \caption{Pictorial depiction of QNMs as poles in the complex $\omega$ plane and of the decay-time shift in theories beyond GR. Different points correspond to different overtones at fixed $\ell$ and $m$. The stability bound establishes the existence of a prohibited region for each shifted mode.}
    \label{fig:qnms}
\end{figure}

In \cite{Melville2401}, a bound on the possible deviations of the QNMs spectrum from GR was conjectured:
\begin{equation} \label{eq:stabbound}
  \Re[\omega_{\rm GR}] \delta \tau \lesssim 1
\end{equation}
where $\omega_{\rm GR}$ is the corresponding mode in GR. This bound was used to constrain the EFT coefficients of irrelevant operators in GREFT up to mass dimension 8. We refer to \eqref{eq:stabbound} as the \textit{stability bound}. The physical interpretation of the bound is the following: black holes in (supposedly) healthy theories of modified gravity are always \textbf{more stable} than those in GR. In this context, \say{more stable} indicates that the QNMs decay more rapidly, allowing the Black hole to relax to a stationary configuration over a shorter timescale. If they were to be less stable than GR (so positive $\delta \tau$), the shift would have to be so small to be \textit{unresolvable}, meaning smaller than $1/\Re[\omega]$. We will return to the notion of resolvability shortly. A pictorial depiction is provided in Figure \ref{fig:qnms}. 

\noindent In \cite{Melville2401} three theoretical justifications to support the conjecture are given:
\begin{enumerate}
    \item the first justification is related to the interplay between the poles of Green's functions and heavy degrees of freedom. The reasoning is borrowed from QFT, where the renormalization of the two point function changes the position of the poles of the propagators. The author argues that since the contribution from heavy degrees of freedom on IR physics always implies a shift of the poles towards negative values in the imaginary ($\omega$) axis, the same should be true for quasinormal modes in GREFT, where the IR action is obtained by integrating out heavy degrees of freedom.
    \item The second one is related to the fact that, since QNMs are wave perturbations, they should be treated on the same footing as scattered gravitational waves. Causality bounds (for example \cite{de_Rham2112}) constrain EFT coefficients requiring propagating waves on curved backgrounds to be asymptotically slower than light. In other words, time advances must not occur in healthy EFT. By association, something analogous is assumed to hold for the decay-time shift.
    \item The third is empirical: in the theories examined in \cite{Melville2401}, the positivity and causality bounds limit, or remove altogether, any positive $\delta\tau$. The author takes this agreement as evidence that \eqref{eq:stabbound} is a further manifestation of causality, and finds that imposing it reproduces the existing constraints and in some cases sharpens them.
\end{enumerate}
The next subsection examines the second argument and, at its end, the first in its most direct form; the third is the subject of Section~\ref{sec:4}.

\subsection{Comments on time advances and resolvability}
Relativistic causality forbids information from propagating outside the
causal cone. In low-energy EFT analyses one commonly adopts an operational diagnostic,
asking whether an EFT-induced time advance is large enough to constitute a
controlled and resolvable effect within the regime of validity of the EFT
and of the semiclassical or eikonal approximation \cite{de_Rham2007, de_Rham2112, Serra2205, Arkani_Hamed2021, Bellazzini2023, Bellazzini2022}. In the case of a probe wave packet, an energy-dependent scattering phase displaces
the packet by
\begin{equation}
    \delta T=\frac{\partial\delta(E)}{\partial E}.
\end{equation}
For a relativistic packet with order-one fractional bandwidth $\Delta\omega\sim\omega_0$, a necessary condition for the shift to produce an order-one displacement
is parametrically
\begin{equation}\label{eq:omegaresolvability}
    \omega_0\,|\delta T|\gtrsim1.
\end{equation}
Quantum uncertainty enters here through the width of the chosen single-particle packet: \eqref{eq:omegaresolvability} is a criterion for an order-one packet displacement. A resolvable time advance is a negative shift satisfying both $\delta T<0$ and \eqref{eq:omegaresolvability}, or equivalently $\omega_0\,\delta T\lesssim-1$. Requiring that the EFT not predict such a resolvable advance therefore
gives
\begin{equation}\label{eq:timeadvance}
    \omega_0\,\delta T\gtrsim-1.
\end{equation}
This condition requires that no negative shift becomes larger than the probe
packet width while the effect remains calculable within the EFT, it is consequently an operational criterion for the controlled diagnosis of an infrared time advance. 

By analogy with the scattering case \eqref{eq:omegaresolvability}, an inequality of this form was
carried over in \cite{Melville2401} from time advances to QNM decay-time
shifts, motivating the criterion
\begin{equation}\label{eq:qnmresolvability}
    \Re[\omega_{\rm GR}] \,|\delta\tau| \gtrsim 1 \, .
 \end{equation}
The analogy is, however, delicate. A QNM is a pole of
the retarded response function, not a wave packet and not a single
quantum. The excited ringdown waveform may admit a semiclassical
description as a coherent gravitational field, but its complex frequency
is a property of the response operator. 

Consider first a quantum reading. The decay time is
inferred from the time dependence of the full waveform, including its
amplitude and phase, rather than from an irreducible arrival-time spread of
one graviton. In an ideal shot-noise-limited description, a ringdown
containing $N_{\rm eff}$ independent units of accessible information, or
measured with waveform signal-to-noise ratio $\rho$, gives parametrically
\begin{equation}
    \frac{\sigma_\tau}{\tau}
    \sim \frac{1}{\sqrt{N_{\rm eff}}}
    \sim \frac{1}{\rho},
\end{equation}
up to order-one factors. We have
\begin{equation}
    \Re[\omega] \,\sigma_\tau
    \sim \frac{ \Re[\omega]\tau}{\rho} .
\end{equation}
A decay-time shift satisfying $|\delta\tau|\ll 1/  \Re[\omega]$ can therefore
be statistically resolvable when $\rho\gg  \Re[\omega]\tau $. The relevant quantity here
is the information about the ringdown parameters that is accessible to the
measurement. This rules out an occupation-independent Heisenberg
justification of \eqref{eq:qnmresolvability}.

Consider next a classical or operational reading. The Fourier
time-bandwidth relation applies to classical fields and to
quantum modes, but it constrains the temporal and spectral shape
of the mode; it does not impose a Fourier lower bound on the error with
which a parameter of a known waveform can be estimated. In an ideal
noiseless and exactly modelled experiment, this estimation error need not
be of order $1/  \Re[\omega]$. In a realistic experiment it is instead
controlled by detector noise, signal-to-noise ratio, observation
protocol, and degeneracies with other waveform parameters. At finite signal-to-noise, a sufficiently small $\delta\tau$ may of course
be difficult to distinguish from the GR waveform. This is a statement
about observability, not about whether the corrected QNM frequency is
well-defined in the underlying theory. The precision can be improved by
increasing the total signal-to-noise, lowering the detector noise, or
jointly analysing many events under a common EFT model. Simply extending
a fixed ringdown to arbitrarily late times does not provide unlimited
information, since the exponentially decaying signal eventually becomes
negligible compared with the noise. Conversely, an EFT calculation does
possess a distinct theoretical accuracy limit: a correction smaller than
the estimated contribution of neglected operators or higher
perturbative orders cannot be robustly attributed to a particular Wilson
coefficient. This is a truncation error of the EFT, not an intrinsic
$1/ \Re[\omega]$ uncertainty in the definition of $\tau$. Laser interferometry makes this point particularly clear.
Advanced LIGO uses $1064\,{\rm nm}$ laser light in $4\,{\rm km}$ arms,
while Virgo has $3\,{\rm km}$ arms. The corresponding optical period is
\begin{equation}
    T_{\rm laser}
    =\frac{\lambda}{c}
    =\frac{2\pi}{\omega_{\rm laser}}
    \simeq 3.5\times10^{-15}\,{\rm s}.
\end{equation}
For illustration, a strain amplitude spectral density
$S_h^{1/2}\sim10^{-23}\,{\rm Hz}^{-1/2}$ in a detector with $L\simeq4\,{\rm km}$ corresponds, up to the precise strain convention and detector response, to a differential-displacement amplitude spectral density
\begin{equation}
    S_{\Delta L}^{1/2}
    \sim L S_h^{1/2}
    \sim4\times10^{-20}\,
        {\rm m\,Hz}^{-1/2},
\end{equation}
or to the equivalent dimensional scale
\begin{equation}
    S_{\Delta t}^{1/2}
    \equiv\frac{S_{\Delta L}^{1/2}}{c}
    \sim1.3\times10^{-28}\,
        {\rm s\,Hz}^{-1/2}.
\end{equation}
Numerically, over a $1\,{\rm Hz}$ reference bandwidth, this scale is
about $4\times10^{-14}$ of one optical period. This comparison should not be interpreted as a direct single-shot measurement of a light-travel-time delay of $10^{-28}\,{\rm s}$. The interferometer measures a differential optical phase, and its sensitivity is accumulated over a finite bandwidth and observation time. Nevertheless, it illustrates the relevant principle: a coherent phase measurement is not limited to resolving shifts of one carrier cycle. In an ideal unsqueezed coherent-state description, the shot-noise-limited phase uncertainty decreases as $1/\sqrt{N_\gamma}$, rather than approaching an occupation-independent floor of order $1/\omega_{\rm laser}$. 

The corresponding lesson for ringdown is that the excited gravitational-wave field is a macroscopic phase-coherent waveform whose parameters are inferred from the full time series. The uncertainty in its decay time is controlled by the gravitational-wave signal-to-noise ratio, parameter correlations, detector noise, and waveform systematics, rather than by an intrinsic one-cycle uncertainty of order $1/ \Re[\omega]$. 

Where does this leave the stability bound? If \eqref{eq:qnmresolvability} does not apply to QNMs, the natural alternative would be to strictly demand
\begin{equation}\label{eq:strictalternative}
    \Re[\omega_{\rm GR}] \delta \tau \leq 0 \, ,
\end{equation}
with no resolvability tolerance whatsoever. This is the version that the heuristic ``heavy physics pushes the poles deeper into the lower half-plane'' would suggest in its most direct form. However, it appears too strong. In the Schwarzschild GREFT expansion considered below (see Section \ref{sec:4}), the six-derivative cubic-curvature, or $\mathbf{Riem}^3$, interaction shifts the
odd- and even-parity decay times with opposite signs. Imposing \eqref{eq:strictalternative} mode by mode in both sectors would therefore force the net $\mathbf{Riem}^3$ coefficient to vanish. It would consequently exclude UV completions whose low-energy pure-gravity description generates a non-zero net $\mathbf{Riem}^3$ coupling\footnote{This is not the case for superstring theories, where supersymmetry forbids that term \cite{GRISARU1986347, GROSS19861}, but it is for bosonic string theory.}. We henceforth do not adopt this monotonicity conjecture. 

We are then left taking the stability bound \eqref{eq:stabbound} on its own terms, having concluded that the first two arguments do not extend straightforwardly to QNMs. The remainder of the paper asks what its content nonetheless implies for the EFT. The answer is developed in Section \ref{sec:4}.

\section{On the resolvability of decay-time shifts}\label{sec:4}
Even taken on its own terms, we find that the stability bound \eqref{eq:stabbound} does not yield nontrivial constraints on the Wilson coefficients within the EFT's regime of validity. Indeed, in the case of a single scale EFT, $\Re[\omega_{\rm GR}]\delta\tau$ reaches $\mathcal{O}(1)$ only at the edge of perturbativity, where predictivity is lost. With \say{single scale EFT} we mean a theory where both the derivative and the operator expansion are weighted by the same scale $\Lambda$. Said differently, the bound is thus automatically satisfied by any predictive EFT extension of GR with a known UV completion, implying that no condition on the coefficients can ever be extracted from it. The stability bound is of no practical use.  For the sake of completeness, we also discuss EFTs where the derivative expansion is weighted by a different scale, $\Lambda_{\nabla}$ and find possible enhancements, though no new constraint is obtained on GREFT. 

Our argument is a scaling one: we track how $\Re[\omega_{\rm GR}]\delta\tau$ depends on the mode frequency and the EFT scales, and ask whether it can reach $\mathcal{O}(1)$ anywhere in the EFT's regime of validity. Since it grows with $\omega$, we push $\omega$ as high as the EFT can accommodate — the eikonal regime — and estimate $\delta\tau$ there via WKB. The estimate requires extending the standard WKB treatment to effective potentials polynomial in $\omega$ and $\ell$, and keeping track of the additional scales that higher-derivative corrections bring in. Readers familiar with the standard analysis can skip to Section \ref{sec:3.2}. 

\subsection{WKB approach for decoupled QNMs equations} \label{sec:3.1}
In GR, small perturbations of a Schwarzschild Black hole obey two decoupled second-order ODEs -- the Regge-Wheeler and Zerilli equations -- in which $\omega$ enters only through the term $\omega^2$, since the Einstein-Hilbert action contains at most two derivatives acting on the metric perturbations. In an EFT, higher-derivative operators introduce additional energy scales -- collectively denoted $\Lambda$ below -- and the master potential acquires arbitrary polynomial dependence on $\omega$ and on the angular momentum $\ell$. We extend the standard WKB treatment to accommodate this. Here we restrict to decoupled equations of motion; the extension to coupled systems is deferred to Appendix \ref{app:A}. 

Consider a master equation of the form
\begin{equation}
    \frac{d^2 \psi}{dr^{2}_*} + V(\Lambda, \omega, \ell; r_*)\,\psi = 0\,,
\end{equation}
where $V$ is polynomial in $\omega$ and $\ell$, and $\Lambda$ denotes the additional scales altogether. Under the assumptions:
\begin{enumerate}
    \item $V(\Lambda, \omega, \ell; r_*\rightarrow \pm \infty) \rightarrow \Tilde{k}_{\pm}(\Lambda, \omega, \ell)$, ensuring wave-like behavior of $\psi$ at infinity and near the horizon;
    \item $\psi(r_* \rightarrow \pm \infty)  \propto e^{\pm i \sqrt{\Tilde{k}_{\pm}} \ r_*}$ (boundary conditions);
    \item the beyond-GR corrections are perturbative, in the sense that they neither alter the qualitative shape of the potential nor destroy the single extremum structure\footnote{The perturbativity assumption can be relaxed: what one really needs is that $-V$ has a maximum at finite $r_*$ and decays monotonically from some $|r_*|$ onward.} \cite{Iyer0687},
\end{enumerate}
the WKB quantization condition at leading order reads\footnote{In \cite{Iyer0687}, the asymptotic behavior of $V$ at infinity is assumed to be $\omega^2$, as in GR. The result, in fact, holds for generic $\omega$-, $\ell$-, and $\Lambda$-dependent asymptotes: the quasinormal-mode quantization formula comes from the vanishing of the incoming wave at the horizon and infinity, and imposes no further requirement on $V$ at $r_*\to+\infty$.}:
\begin{equation} \label{eq:quantcond}
   \frac{i\,V(r_{\text{max}},\omega)}{\sqrt{2\,V^{(2)}(r_{\text{max}},\omega)}} = n + \tfrac{1}{2}\,,\qquad n\in\mathbb{N}_0\,,
\end{equation}
where
\begin{equation}
    V^{(k)}(r_{\text{max}}, \omega) \equiv \left. \frac{\partial^k V(\Lambda, \omega, \ell; r_*)}{\partial r_*^k} \right|_{r_{\text{max}}}\,,
\end{equation}
and $r_{\text{max}}$ is the location of the (single) minimum of $V$:
\begin{equation}
    \left. \frac{\partial V(\Lambda, \omega, \ell; r_*)}{\partial r_*} \right|_{r_{\text{max}}} = 0\,.
\end{equation}
The integer $n$ is the overtone number; we set $n=0$ throughout. 

We decompose the effective potential as
\begin{equation}
    V(\Lambda, \omega, \ell; r_*) = \omega^2 - V_{\text{GR}}(\ell; r_*) + \delta V(\Lambda, \omega, \ell; r_*)\,,
\end{equation}
with $\delta V$ the EFT correction, polynomial in $\omega$ and $\ell$ as before. The QNM frequencies in the modified theory are then obtained by solving \eqref{eq:quantcond} for $\omega$. Assuming the EFT corrections to be small compared to the GR contribution, both the QNM frequency and $r_{\text{max}}$ shift perturbatively away from their GR values, $\omega^{\text{EFT}} = \omega_{\rm GR} + \delta\omega$ and $r_{\text{max}}^{\text{EFT}} = r_{\text{max}}^{\text{GR}} + \delta r_{\text{max}}$, so that to first order
\begin{equation}
    V^{(k)}\left(r_{\text{max}}^{\text{EFT}}, \omega^{\text{EFT}}\right) \approx -V^{(k)}_{\text{GR}}\left(r_{\text{max}}^{\text{GR}}\right) - V^{(k+1)}_{\text{GR}}\left(r_{\text{max}}^{\text{GR}}\right) \delta r_{\text{max}} + \delta V^{(k)}\left(r_{\text{max}}^{\text{GR}}, \omega_{\rm GR}  \right) 
\end{equation}
for $k\geq 1$. Using the definition of $r_{\text{max}}$, we also obtain the relation: 
\begin{equation}
    \delta r_{\text{max}} =  \left.\frac{\delta V^{(1)}}{V_{\text{GR}}^{(2)}}\right|_{r_{\text{max}}^{\text{GR}}}
\end{equation}
that allows us to write $r_{\text{max}}^{\text{EFT}}$ as a function of GR quantities only. Furthermore, it is possible to extract $\omega^{\text{EFT}}$ without solving an arbitrary polynomial equation, but only through \eqref{eq:quantcond}, as follows: 
\begin{equation}
   \begin{split}
       \left( \omega^{\text{EFT}}\right)^2 \approx & V_{\text{GR}} - \delta V  + \frac{1}{2} \sqrt{2 V^{(2)}_{\text{GR}}} \times 
        \left( 1 +\frac{V^{(3)}_{\text{GR}}}{\left(V^{(2)}_{\text{GR}}\right)^2} \delta V^{(1)} - \frac{\delta V^{(2)}}{V^{(2)}_{\text{GR}}}  \right)^{1/2}
   \end{split}
\end{equation}
where the functions are evaluated at $r_{\text{max}}^{\text{GR}}$. Since $V_{\text{GR}}$ has a maximum there, $V^{(2)}_{\text{GR}}<0$, and the square root must be taken on the branch
\begin{equation}
    \sqrt{2V^{(2)}_{\text{GR}}}=-iA\,,\qquad A\equiv\sqrt{-2V^{(2)}_{\text{GR}}}>0\,.
\end{equation}
At first order, $2 \omega_{\rm GR} \delta \omega \approx \Delta V$, where
\begin{equation}
      \Delta V := \sqrt{2 V^{(2)}_{\text{GR}}}\left(  -\frac{\delta V}{\sqrt{2 V^{(2)}_{\text{GR}}}} + \frac{V^{(3)}_{\text{GR}}}{4\left(V^{(2)}_{\text{GR}}\right)^2} \delta V^{(1)} - \frac{\delta V^{(2)} }{4V^{(2)}_{\text{GR}}} \right).
\end{equation}
Selecting the imaginary part,
\begin{equation}\label{eq:imaginarycorrecction}
       \Im[\delta \omega] \approx \frac{ \Re[\omega_{\rm GR}] \Im[\Delta V]-\Im[\omega_{\rm GR}] \Re[\Delta V]}{2\abs{\omega_{\rm GR}}^2} 
\end{equation}
we can then compute the decaying time difference of quasinormal modes in GR and modified theories of gravity as:
\begin{equation}
    \delta \tau  = \frac{2\pi}{\Im[\omega_{\rm GR}]}-\frac{2\pi}{\Im[\omega^{\text{EFT}}]} \approx 2\pi \frac{\Im[\delta \omega]}{\Im[\omega_{\rm GR}]^2} \implies \Re[\omega_{\rm GR}] \delta \tau \approx 2\pi \frac{\Re[\omega_{\rm GR}]}{\Im[\omega_{\rm GR}]^2} \times \Im[\delta \omega]
\end{equation}
Crucially, the decay-time shift can be expressed solely in terms of GR quantities and the shift in the potential. The WKB approximation assumes $\omega$ to be large enough. This assumption is equivalent to imposing
\begin{equation}
    \Re[\omega_{\rm GR}] r_s \gg 1 \qquad \ell  \gg 1 \qquad \text{and} \qquad \frac{\Re[\omega_{\rm GR}] r_s}{\ell} \rightarrow \text{const.}, 
\end{equation}
some remarks on this point: 
\begin{itemize}
    \item since we assume that $\delta V$ is polynomial in $\omega$ and $\ell$, \eqref{eq:imaginarycorrecction} is likewise polynomial in $\ell$ and $\omega_{\rm GR}$;
    \item at higher orders in the WKB expansion, \eqref{eq:imaginarycorrecction} would receive additional contributions that remain linear in $\delta V$ and its derivatives, but are subleading in the large-$\ell$/$\omega_{\rm GR}$ expansion. Since in the following we focus on the large-angular-momentum regime, these higher-order contributions need not be included. In other words, the leading-order eikonal approximation adopted here becomes accurate precisely in the regime of interest;
    \item the range of $\omega^{\text{EFT}}$, and consequently that of $\omega_{\rm GR}$ and $\ell/r_s$, is ultimately limited by the cutoff of the theory. This is the crucial point that will be discussed in detail in the next section.
\end{itemize} 
The scaling we will be using in the subsequent sections is:
\begin{equation} \label{eq:resolvabilityeq}
    \boxed{\Re[\omega_{\rm GR}] \delta \tau \sim  \frac{\Delta V}{\Im[\omega_{\rm GR}]\Re[\omega_{\rm GR}]} }
\end{equation}
where both the real and imaginary part of $\Delta V$ contribute. This result will serve as the basis for the following analysis. 

\subsection{Corrections resolvability in GREFT} \label{sec:3.2}
The effective way of describing a graviton degree of freedom alone at low energies, when heavy degrees of freedom are integrated out is\footnote{We are discussing the case in which the high energy physics is purely gravitational, implying that the low energy description does not include any additional coupling constant. Here gravity is \textit{composite} at high energy, as in String Theory. For reasons that are provided in details in the Appendix \ref{app: B} within the context of scalar-tensor theories, but equivalently valid in GREFT, the EFT corrections in the composite scenario are more enhanced compared to the elementary case and for the purpose of our claim, an enhancement is what we need.}:
\begin{equation} \label{eq:NDALagrangianGREFT}
    \mathcal{L} = \frac{1}{2}M_P^2 R + M_P^2 \Lambda^2 \left(\hat{\mathcal{L}}_{\text{tree}}\left( \frac{\nabla}{\Lambda_{\nabla}} , \frac{\mathbf{Riem}}{\Lambda^2} \right) + \frac{\Lambda^2}{16 \pi^2 M_P^2} \hat{\mathcal{L}}_{\text{1-loop}}\left( \frac{\nabla}{\Lambda_{\nabla}} , \frac{\mathbf{Riem}}{\Lambda^2} \right) + \cdots \right)
\end{equation}
where we also allow the possibility that the derivative and field power counting are weighted by different scales. The most generic IR theory includes the whole tower of effective operators, and, when $\Lambda \sim \Lambda_{\nabla}$ its cutoff coincides with $\Lambda$. Restricting to selected choices of tree level operators, for example motivated by symmetry, result in a cut-off that depends on $\Lambda$, $\Lambda_{\nabla}$, $M_P$ and, since we aim to specialize the discussion to spherically symmetric black holes, an additional scale, meaning $r_s$. The cut-off can coincide with the strong coupling scale $\Lambda_{\text{s.c.}}$, meaning the highest possible scale at which new physics appears, or can be parametrically lower, depending either on consistency considerations (including causality, positivity etc.) or the appearance of \say{unexpected} degrees of freedom which were not taken into account to begin with. 

Assuming the parametric hierarchy:
\begin{equation} \label{eq:hierarchy}
    4 \pi M_P \gg \left\{\Lambda, \Lambda_{\nabla} \right\} \gg \frac{1}{r_s}, 
\end{equation}
the loop corrections are suppressed relative to the tree-level operators. We henceforth focus on tree-level contributions, keeping in mind that all operators are generically generated at loop level unless forbidden by symmetries. The hierarchy between $\Lambda$ and $\Lambda_{\nabla}$ is not arbitrary. In particular, the only two possibilities are $\Lambda \sim \Lambda_{\nabla}$ and $\Lambda \ll \Lambda_{\nabla}$. This is due to the commutation relation $[\nabla, \nabla] \sim \textbf{Riem}$, implying that couples of derivatives, always generate powers of the Riemann tensor. If the Riemann tensor was suppresses by a scale larger than the one suppressing covariant derivatives, less suppressed curvature tensors would always be generated by covariant derivatives, for instance: 
\begin{equation}
    \left(\frac{\nabla}{\Lambda_{\nabla}}\right)^2 \left(\frac{\textbf{Riem}}{\Lambda^2} \right)^4 \rightarrow \frac{\textbf{Riem}^5}{\Lambda^8\Lambda^2_{\nabla}} \gg \left(\frac{\textbf{Riem}}{\Lambda^2} \right)^5.
\end{equation}
As a consequence, every dominant contribution, even in the Riemann expansion, would be weighted by the scale of the derivative expansion\footnote{The bound is one-sided because the identity $[\nabla,\nabla]\sim\mathbf{Riem}$
runs in a single direction: pairs of covariant derivatives generate curvature, but the
Riemann tensor is the field strength of the connection and cannot be traded for covariant
derivatives of any covariant object just as $[D,D]\sim F$ yields a field strength that is not itself the covariant derivative of
anything covariant. Hence $\Lambda \ll \Lambda_\nabla$ stays consistent, the curvature
regenerated by derivative pairs, $\mathbf{Riem}/\Lambda_\nabla^2$, being more suppressed
than the nominal $\mathbf{Riem}/\Lambda^2$, whereas $\Lambda_\nabla \ll \Lambda$ does not.}.

The relevant tensors can be written, suppressing the indices for simplicity, as: 
\begin{equation}
    \textbf{Riem} \sim \partial^2  g \qquad R \sim  g^2 \partial^2  g \qquad \text{where} \qquad g \sim \Bar{g} + \frac{h}{M_P}
\end{equation}
where we denote quantities evaluated on the background as barred. In the case of a modified Schwarzschild background: 
\begin{equation}
  \Bar{g} = 1 - \frac{r_s}{r} + \mathcal{O}\left(\frac{1}{r_s\Lambda}\right)\qquad  \overline{\textbf{Riem}} = \partial^2 \Bar{g} \sim \frac{r_s}{r^3} + \mathcal{O}\left( \frac{1}{r_s\Lambda}\right)
\end{equation} 
The equations of motion can be found expanding the generic Lagrangian \eqref{eq:NDALagrangianGREFT} to second order in $h_{\mu \nu}$ and taking the variation. Making use of the static nature and spherical symmetry of the background solutions we are considering, we can decompose the metric perturbations as:
\begin{equation}
    h(t, r, \Omega)  \rightarrow e^{- i \omega t} Y_{\ell m}(\Omega) h(r)
\end{equation}
Generically, discarding the higher derivatives terms in $r$, that we treat as perturbative corrections, we obtain a set of ten second order ODEs. After choosing a gauge, the system still includes all the auxiliary non-propagating fields, but some of the equations will become constraints. By choosing the correct linear combinations, the dynamical equations will reduce to a set of two coupled ODEs, for which it can be proven \cite{Hui2311} that the first derivative terms are eliminated by a field redefinition, together with the mixed second order derivatives. Making use of tortoise coordinates $r\rightarrow r_*$, the result is, generically: 
\begin{equation} \label{eq:mastereqGREFT}
    \frac{d^2 \psi_{\pm}}{d r^{*2}} + \left(\omega^2 - V_{\text{GR}}^{\pm}(\ell; r_*) \right) \psi_{\pm} + \sum_{s=\pm}  \delta V_{s\pm}(\Lambda, \Lambda_{\nabla}, \omega, \ell; r_*)\psi_s = 0 
\end{equation}
These are the master equations for the two propagating gravitational degrees of freedom $\psi_{\pm}$. The usefulness of writing them in this form is to make clear the power counting in all possible EFT extensions of GR on the Black hole's background. If the action does not include any parity breaking term, the even and odd perturbations decouple and can be studied easily using the WKB approach. In particular, we have: 
\begin{equation} \label{eq:GREFTEOM}
\begin{split}
   \frac{d^2\psi_{\pm}}{dr^{*2}} +\left(\omega^2 - V_{\text{GR}}^{\pm}(\ell; r_*) + \delta V_{\pm \pm}(\Lambda, \Lambda_{\nabla},\omega, \ell; r_*) \right)  \psi_{\pm} = 0 
\end{split}
\end{equation}
We will estimate the contribution to $\delta V$, and therefore to the decay-time shift \eqref{eq:resolvabilityeq} generated by a generic operator in the EFT expansion.   

\subsubsection{Validity scales}
Since our aim is to enhance the decay-time shift as much as possible, we should evaluate at which energy scale the EFT breaks down. There are two relevant scales to focus on: 1. the scale at which perturbativity of the equations of motion is not valid anymore\footnote{Notice that the scale at which physical ghosts appear is accounted into the EOM perturbativity scale, meaning that ghosts always appear at a higher scale than the one at which the perturbativity of EOM breaks down.} $\Lambda_{\text{pert}}$, 2. the scale at which the theory becomes strongly coupled $\Lambda_{\text{s.c.}}$. 

Referring to Appendix \ref{app: B} for details, the summary for a generic EFT is provided in Table \ref{tab:1}.
\begin{table}[h!]
    \centering
    \renewcommand{\arraystretch}{1.3}
    \begin{tabular}{|c|c|c|c|}
  \hline
  \multicolumn{2}{|c|}{Scenario} & Strong coupling & EOM perturbativity \\ 
  \hline
  \multirow{2}{*}{$\Lambda_{\nabla} \gg \Lambda$} & $\Lambda_{\nabla} \gg r_s\Lambda^2$ & $\Lambda_{\nabla}$ (not $\Lambda$) & $\Lambda (r_s\Lambda)^2$ \\ 
   & $\Lambda_{\nabla} \lesssim r_s\Lambda^2$ & $\Lambda_{\nabla}$ (not $\Lambda$) & $\Lambda(r_s\Lambda_{\nabla})$ \\ 
  \hline
  \multicolumn{2}{|c|}{$\Lambda_{\nabla} \sim \Lambda$} & $\Lambda_{\nabla} \sim \Lambda$ & $r_s\Lambda_{\nabla}^2\sim r_s \Lambda^2$ \\
  \hline
        \end{tabular}
        \caption{The relation between the two scales is determined by the fact that $r_s \Lambda \gg 1$.}
        \label{tab:1}
\end{table}
In the single-scale case the perturbativity scale lies parametrically above the strong-coupling scale, where new physics should appear. This is consistent, because the strong-coupling scale bounds local Lorentz invariants, such as the center-of-mass energy of graviton scattering, while a quasinormal mode is a property of a single fluctuation on the black-hole background. For such a fluctuation the frequency cannot enter the EFT expansion on its own: once the leading equations of motion are used, local operators can only contract the wave vector with the background curvature. The invariants that grow with the frequency then measure, in effect, the center-of-mass energy of the wave scattering off the background, which is much smaller than the frequency and reaches the scale of new physics only at the perturbativity scale, where a generic EFT loses predictivity \cite{de_Rham2007, Chen2112}. Processes involving several independent fluctuations remain bounded by the strong-coupling scale instead. Taking the perturbativity scale, as in \cite{Melville2401}, is in any case the choice most favorable to the stability bound: a lower cutoff would only strengthen our conclusions. 

One might think that it is possible to select specific IR realizations with a huge enhancement of the EOM perturbativity scale. An example is an EFT that includes \textbf{only} the operator $\textbf{Riem}^3$, and no other term in the tree level EFT expansion. Nonetheless, this fails when including the first loop corrections. The most stringent bound on the perturbativity scale stems from high derivative operators, for which the Planck suppression is immaterial: even for IR theories with a classical enhancement of the perturbativity scale, quantum corrections bring it back down to $\Lambda(r_s\Lambda)$. 

\subsubsection{Decay-time shift resolvability} 
As it should be clear from the previous discussion, if we consider EFT corrections with a chosen number of derivatives $2\J$ and powers of the Riemann tensor $\R$, the dominant contribution in the potential will be given by:
\begin{equation}\label{eq:potshift}
    \delta V(\Lambda_{\nabla}, \Lambda, \omega, \ell; r_*) \supset \Lambda^2\left(\frac{1}{\Lambda_{\nabla}}\right)^{2\J} \left( \frac{1}{\Lambda^2}\right)^{\R} \kappa^{n+4} \times G\left(r_*, r_s, \left\{ b_i \right\} \right) 
\end{equation}
where $\kappa$ denotes either $\omega$ or $\ell/r_s$, $G$ is a function of $r_*$, $r_s$ and the EFT parameters $\left\{ b_i \right\}$ of mass dimension $[G]= 2(\J + \R-2)-n$ , and $n$ is a positive integer bounded from above (see Appendix \ref{app: B} for details). This contribution is present unless some selection rules precludes it from appearing, in such a case only lower powers in $\kappa$ will be present, like $\kappa^{n+3}$. By use of the eikonal formula for quasinormal modes frequencies \eqref{eq:imaginarycorrecction}, observing that the RHS frequencies are the GR ones, we obtain: 
\begin{equation}
   \Im[\delta \omega] \sim \frac{1}{r_s} \left(\frac{1}{r_s \Lambda_{\nabla}}\right)^{2\J} \left( \frac{1}{r_s \Lambda} \right)^{2\R-2} \ell^{n+2} \times F\left(\left\{b_i\right\}\right)
\end{equation}
where now $F$ is a dimensionless function of the EFT coefficients only and it's of order one. Using \eqref{eq:resolvabilityeq}, the dominant contributions at large $\ell$ are\footnote{See Appendix \ref{app: B} for the condition on $n$ in the $\R=3$ case.}: 
\begin{equation} \label{eq:eikonalshift}
    \Re[\omega_{\rm GR} ] \delta \tau \sim  H(\left\{ b_i\right\}) \times
    \begin{cases}
        \left( \frac{1}{r_s\Lambda}\right)^4 \ell   & \qquad  \text{for} \quad \R = 3, \J = 0 \\
         \left(\frac{1}{r_s\Lambda_{\nabla} }\right)^{2\J} \left(\frac{1}{r_s\Lambda }\right)^{2\R-2}  \ell^{\R+\J-1} & \qquad \text{for} \quad \J > \R-4, \ \R\geq 4 \\ 
          \left(\frac{1}{r_s\Lambda_{\nabla}}\right)^{2\J} \left(\frac{1}{r_s\Lambda}\right)^{2\R-2}   \ell^{2\J+3} 
           & \qquad \text{for} \quad \J \leq \R-4, \ \R \geq 4
    \end{cases}
\end{equation}
where $H$ is a different dimensionless function of order one in the expansion coefficients. Before providing some examples, notice that, since $\ell_{\text{max}}\gtrsim (r_s\Lambda)^2$, from Table \ref{tab:1}, when $\R<\J+4$, the higher $\R$, the larger the shift. On the other hand, when $\R\geq\J+4$, it is clear that the larger $\R$, the smaller the shift. As a result, the largest decay-time shift is obtained for
\begin{equation}\label{eq: maxshift}
   \left. \Re[\omega_{\rm GR} ] \delta \tau \right|_{\text{max}}  \sim \left( \frac{\ell}{(r_s\Lambda)(r_s\Lambda_{\nabla})} \right)^{2\J} \left(\frac{\ell}{(r_s\Lambda)^2}\right)^3  \times H
\end{equation}
To assess the validity of the estimate \eqref{eq:eikonalshift}, we compare it with the explicit examples provided by \cite{Melville2401}. For $\J=0$ we find at mass dimension $6$ and $8$, for $\Lambda_{\nabla} \sim \Lambda$:
\begin{equation} \label{eq:p3p4cases}
\Re[\omega_{\rm GR} ] \delta \tau \sim  
    \begin{cases}
        \left(\frac{1}{r_s \Lambda}\right)^4 \ell \times H\left(b_1\right) &\qquad  \R=3 \\ 
          \left(\frac{1}{r_s \Lambda}\right)^{6} \ell^3 \times H\left(c_1, c_2\right) & \qquad  \R=4 
    \end{cases}
\end{equation}
This is the same behavior of Equation (3.44)-(3.45) in \cite{Melville2401} with\footnote{Although these functions can attain values much larger than one for order-one coefficients, there is no scale or expansion parameter that separates them from unity. Our interpretation is that, despite their numerically large values, these quantities should be regarded as parametrically of order unity.}:
\begin{equation}
\begin{split}
    & H\left(c_1, c_2\right) \approx - 1.9\times 10^3 \ c_2  \\ 
    & H\left(c_1, c_2\right) \approx - 4.7 \times 10^2  \ c_1 
\end{split} \qquad \begin{split}
    & H\left(b_1\right) =  320 \ b_1  \qquad \text{odd sector} \\ 
    & H\left(b_1\right) = - 320 \ b_1 \qquad  \text{even sector}
\end{split}
\end{equation}

\noindent The question, at this point, is whether the decay-time shift is resolvable or not provided that the expansion coefficients $\left\{ b_i \right\}$ are all of order one.  As should be clear by now, resolvability is related to maximum value of $\ell/r_s$ within control\footnote{One may wonder about the case in which $r_s \Lambda $ or $ r_s \Lambda_{\nabla}$ are forced to be small, thereby suppressing the denominator in \eqref{eq:eikonalshift}. While this may render the decay-time shifts possibly resolvable, it would at the same time render the EFT essentially useless, since black holes themselves would no longer be resolved within the EFT.}. The answer follows immediately from \eqref{eq: maxshift}. 
\begin{table}[h]
    \centering
    \renewcommand{\arraystretch}{1.6}
    \resizebox{\textwidth}{!}{%
    \begin{tabular}{|cc|c|c|c|c|}
    \hline
     \multicolumn{2}{|c|}{Case} 
    & $\ell_{\text{max}}$ 
    & $\left.\Re[\omega^{\mathrm{GR}}]\delta \tau \right|_{\text{max}}$ 
    & Resolvable 
    & Bound \\ 
    \hline

     \multirow{2}{*}{$\Lambda_{\nabla} \gg \Lambda$} 
    & $\Lambda_{\nabla} \gg r_s\Lambda^2$ 
    & $(r_s\Lambda)^3$
    & $\propto (r_s\Lambda)^3$
    & Yes 
    &  $\equiv$ positivity \\

    & $\Lambda_{\nabla} \lesssim r_s\Lambda^2$
    & $(r_s\Lambda_{\nabla})(r_s\Lambda)$
    & $\propto \left(\frac{\Lambda_{\nabla}}{\Lambda}\right)^3$ 
    & Yes
    &  $\equiv$ positivity \\

     \multicolumn{2}{|c|}{$\Lambda_{\nabla} \sim \Lambda$} 
    & $(r_s\Lambda)^2$
    & $\mathcal{O}(1)$ 
    & Not parametrically 
    & Not useful \\
    
    \hline  
\end{tabular}}
    \caption{Mostly enhanced values of decay-time shifts. Equivalent to positivity means that there is no possibility of extracting any new parametric constraint compared to positivity bounds. Irrespectively, the stability bound is blind to anything but $(\R, \J) = (4, 0)$.}
    \label{tab:result1}
\end{table}

The two $\Lambda_{\nabla}\gg\Lambda$ rows of Table~\ref{tab:result1}
deserve comment, since there the shift exceeds unity \emph{within} the regime of
validity. Writing $\ell_{\max}$ for the corresponding entry of the table, in the
window $(r_s\Lambda)^2\ll\ell\ll\ell_{\max}$ the theory is genuinely predictive:
by \eqref{eq: maxshift}, the shift is dominated by the two
$(\mathrm{J},\mathrm{R})=(0,4)$ operators, higher curvature powers being
suppressed by
\begin{equation}
    \frac{\delta\tau_{\mathrm{J}}}{\delta\tau_{0}}
    = \left(\frac{\ell}{(r_s\Lambda_{\nabla})(r_s\Lambda)}\right)^{2\mathrm{J}}\,,
\end{equation}
so that each extra derivative pair is suppressed by precisely the scale at which perturbativity of the equations of motion breaks down: the statements ``the derivative tower takes over'' and ``the EFT ceases to be predictive'' coincide, as they must. Inside the window, the comparison with Eq.~(3.44) of \cite{Melville2401} can be taken over directly: imposing \eqref{eq:stabbound} in the window is equivalent to $c_1\geq0$ and $c_2\geq0$, up to corrections suppressed by $(r_s\Lambda)^6/\ell_{\max}^3$: where the bound is resolvable, it is redundant. This is the precise sense in which it ``tracks'' the positivity bounds. 

Note that no ultraviolet completion of the two-scale power counting is known: tree-level string amplitudes generate $\Lambda_{\nabla}\sim\Lambda\sim M_s$. More compellingly, low energy positivity can be interpreted as imposing a constraint on the hierarchy between the two scales. Focusing on the operator $\R=4, \J=0$, the bound\footnote{This is valid in the reasonable assumption that the cut-off coincides with the mass of the lightest higher spin particle appearing in the spectrum. The bound relies on assuming that the infrared logs are at most an order ten factor. } $\alpha_4 \lesssim 1/\Lambda_{\text{s.c.}}^6$ derived in \cite{Caron-Huot:2022ugt} together with the power counting \eqref{eq:NDALagrangianGREFT} and the identification $\Lambda_{\text{s.c.}} \equiv \Lambda_{\nabla}$ established above, implies $\Lambda_{\nabla} \lesssim \Lambda$. Combining this condition with the redundancy associated with commuting covariant derivatives, one finds that, if positivity is taken to be a valid low-energy consistency criterion, the only sensible IR realizations are the ones respecting $\Lambda_{\nabla} \sim \Lambda$\footnote{This condition forces the parameter $\epsilon_{\omega}$ used in \cite{Melville2401}, to be $\sim 1$.}. 

\noindent From this, we draw a clear conclusion:
\begin{quote}
    \textit{In a generic EFT with a single scale, $\Lambda_{\nabla}\sim\Lambda$
    --- the only power counting with a known ultraviolet completion, and admitted by positivity --- the
    decay-time shift is \textbf{never parametrically resolvable}.}
\end{quote}
We stress that for $\Lambda_{\nabla}\sim\Lambda$ the $\mathcal{O}(1)$ result, apart from being possible only for a subset of irrelevant operators, is obtained when perturbativity breaks down: whenever $\ell$ reaches its maximal value, the infinite tower of operators with $4\leq \mathrm{R}=\mathrm{J}+4$ contributes at the same order, and the theory has no predictive power there. Any reasonable assumption on $\ell$ implies a much more suppressed result. For $\Lambda_{\nabla}\gg\Lambda$ the same degeneracy is postponed to the upper edge of the window --- and for $\Lambda_{\nabla}\gtrsim r_s\Lambda^2$ it does not occur at all, the window closing instead through the dimension-8 correction itself. 

\subsection{Corrections resolvability in scalar-tensor theories}\label{scalar-tensor}
We repeat the same considerations in the case of shift-symmetric scalar-tensor theories. The low-energy dynamics of the graviton and scalar degrees of freedom are described by the Lagrangian:
\begin{equation}
    \mathcal{L} = \frac{1}{2} \hat{M}_P^2 R + \frac{\Lambda^4}{\chi^2}\hat{\mathcal{L}}_{\text{tree}}\left(\frac{\nabla}{\Lambda_{\nabla}}, \frac{\mathbf{Riem}}{\Lambda^2}, \frac{\chi \partial \phi}{\Lambda^2} \right) + \cdots
\end{equation}
where $\chi$ is the coupling constant associated with the scalar degree of freedom, and $f:=\Lambda/\chi$ is the respective decay constant. By dimensional analysis, the maximum background value that the field can acquire is:
\begin{equation}
    \partial \bar{\phi} \sim \frac{\Lambda}{r_s \chi}\left( 1 + \mathcal{O}\left(\frac{1}{r_s\Lambda}\right)\right).
\end{equation}
Moreover, since we want the background to be a perturbative modification of Schwarzschild, we would expect the field background to disappear for $r_s \Lambda \gg 1$. As a consequence, the actual dominant contribution of the background field is: 
\begin{equation}
    \partial \bar{\phi} \sim \frac{1}{r_s^2  \chi}\left( 1 + \mathcal{O}\left(\frac{1}{r_s \Lambda} \right)\right)
\end{equation}
which is compatible with the results in \cite{Serra2205}. We denote perturbations of the scalar field as $\varphi$. It can be shown (Appendix \ref{app: B}) that the validity scales of this theory are equivalent to the ones found for GREFT when $f \sim \hat M_P$. 

\subsubsection{Decay-time shift resolvability}

Some examples of master equations for scalar-tensor theories are provided, for instance, in \cite{Tattersall1711}. To use the WKB analysis for coupled systems, we should focus on the diagonal elements in the potential matrix (Appendix \ref{app:A}). While the corrections to the potential in the graviton sector are completely equivalent to \eqref{eq:potshift}, up to a factor $(f/\hat{M}_P)^2$, the scalar sector gets corrected as:
\begin{equation}
    \begin{split}
        &  \delta V_{\varphi}(\Lambda_{\nabla}, \Lambda, \omega, \ell; r_*) \supset \Lambda^4 \left(\frac{1}{\Lambda_{\nabla}}\right)^{\J} \left( \frac{1}{\Lambda^2}\right)^{\R}\left( \frac{1}{\Lambda^2}\right)^{\P} \kappa^{n+2} \times G_{\varphi}\left(r_*, r_s, \left\{ b_i \right\} \right)
    \end{split}
\end{equation}
where $\P$ denotes the number of scalar field operators. Using the WKB formulas, we obtain that the mostly enhanced contributions are:
\begin{equation} \label{eq:eikonalshiftST}
    \Re[\omega_{\rm GR} ] \delta \tau_{\varphi} \sim  
    \begin{cases}
         \left(\frac{1}{r_s\Lambda_{\nabla}}\right)^{\J} \left(\frac{1}{r_s\Lambda }\right)^{2\R+ 2\P-4}  \ell^{\R+(\J+ \P)/2-1}\times H\left(\left\{b_i \right\}\right) & \qquad \text{for} \quad \J > 2\R+ \P-4, \ \P\geq 2 \\ 
          \left(\frac{1}{r_s\Lambda_{\nabla}}\right)^{\J} \left(\frac{1}{r_s\Lambda}\right)^{2\R+ 2\P-4}   \ell^{\J+1}\times H\left(\left\{b_i\right\}\right) 
           & \qquad \text{for} \quad \J \leq 2\R+ \P-4, \ \P \geq 2
    \end{cases}
\end{equation}
Following the same reasoning as in GREFT, one readily finds that in order to maximize the decay-time shift, we need $\R = \frac{\J-\P+4}{2}$, and the minimum value of $\P$ allowed, meaning $\P=2$. This implies that the family of dominant operators yields: 
\begin{equation}
    \Re[\omega_{\rm GR} ] \delta \tau_{\varphi} \sim  \left(\frac{\ell
    }{(r_s\Lambda_{\nabla})(r_s\Lambda )}\right)^{\J} \frac{\ell}{(r_s\Lambda)^2} \times H \qquad  \J \ \text{even}
\end{equation}
\begin{table}[h]
    \centering
    \renewcommand{\arraystretch}{1.7}
    \resizebox{\textwidth}{!}{%
    \begin{tabular}{|c|cc|c|c|c|c|}
    \hline
    
    & \multicolumn{2}{c|}{Case} 
    & $\ell_{\text{max}}$ 
    & $\left. \Re[\omega^{\mathrm{GR}}]\delta \tau \right|_{\text{max}}$ 
    & Resolvable 
    & Bound \\ 
    \hline
    
    \multirow{3}{*}{
        \begin{tabular}{@{}c@{}}
            \textbf{$h$ sector}
        \end{tabular}
    }
    & \multirow{2}{*}{$\Lambda_{\nabla} \gg \Lambda$} 
    & $\Lambda_{\nabla} \gg \Lambda \frac{\hat{M}_P}{f} (r_s\Lambda)$ 
    & $\frac{\hat{M}_P}{f}(r_s\Lambda)^3$
    & $\propto \frac{\hat{M}_P}{f} (r_s\Lambda)^3$
    & Yes 
    & $\equiv$ positivity \\ 
    
    &  
    & $\Lambda_{\nabla} \lesssim \Lambda \frac{\hat{M}_P}{f} (r_s\Lambda)$
    & $(r_s\Lambda_{\nabla})(r_s\Lambda)$
    & $\propto \frac{f^2}{\hat{M}_P^2} \left(\frac{\Lambda_{\nabla}}{\Lambda}\right)^3$ 
    & Possibly
    & $\equiv$ positivity \\

    & \multicolumn{2}{c|}{$\Lambda_{\nabla} \sim \Lambda$} 
    & $(r_s\Lambda)^2$
    & $\mathcal{O}(1)$ 
    & Not parametrically 
    & Not useful \\
    
    \hline
    \multirow{3}{*}{
        \begin{tabular}{@{}c@{}}
          \textbf{$\mathbf{\varphi}$ sector}
        \end{tabular}
    }
    & \multirow{2}{*}{$\Lambda_{\nabla} \gg \Lambda$} 
    & $\Lambda_{\nabla} \gg \Lambda \frac{\hat{M}_P}{f} (r_s\Lambda)$ 
    & $\frac{\hat{M}_P}{f} (r_s\Lambda)^3$
    & $\left(\frac{\hat{M}_P}{f}\right)^3 \frac{(r_s\Lambda)^5}{(r_s\Lambda_{\nabla})^2}$
    & Yes
    & Useful\tablefootnote{With two caveats: first that if positivity is assumed as a consistency condition, this case is excluded altogether, and second, that this is true only for $r_s \Lambda_{\nabla} \ll (\hat M_P /f )^{3/2} (r_s \Lambda)^{5/2}$. }\\ 
    
    &  
    & $\Lambda_{\nabla} \lesssim \Lambda \frac{\hat{M}_P}{f} (r_s\Lambda)$
    & $(r_s\Lambda_{\nabla})(r_s\Lambda)$
    & $\propto \left(\frac{\Lambda_{\nabla}}{\Lambda}\right)$ 
    & Yes
    & Useful \\
    
    & \multicolumn{2}{c|}{$\Lambda_{\nabla} \sim \Lambda$} 
    & $(r_s\Lambda)^2$
    & $\mathcal{O}(1)$ 
    & Not parametrically
    & Not useful \\
    \hline        
\end{tabular}}
    \caption{Maximum decay time shifts in scalar-tensor theories.}
    \label{tab:3}
\end{table}

The conclusions largely overlap with those obtained for GREFT, with the only caveat that, when $\ell_{\text{max}}\sim (r_s\Lambda)(r_s\Lambda_{\nabla})$, the resulting shift in the $h$ sector may remain unresolvable. Concerning the $\varphi$ sector, the would-be-dominant operators, meaning the ones with $\J=0$, are redundant, and can be eliminated by a field redefinition. The operator tower in the window $(r_s\Lambda)^2\ll\ell\ll\ell_{\max}$ is therefore dominated by the $\J=2$ operators, corresponding to $(\R,\P,\J)=(2,2,2)$. Consequently, the stability bound could, in principle, provide a probe of the consistency of this specific operator. However, also in scalar-tensor theories, positivity \cite{Caron-Huot:2022ugt, Serra2205} constrains the hierarchy between the two scales $\Lambda_{\nabla}$ and $\Lambda$. In particular, the bounds associated with the Gauss--Bonnet and Chern--Simons operators imply $\Lambda_{\nabla} \lesssim \left( \frac{\hat{M}_P}{f}\right)^{1/4} \Lambda$, which severely restricts the applicability of the stability bound, even in the single case in which it could otherwise have been useful. Irrespectively, partial bounds on those operators' coefficients are already present in \cite{hong2023causalityboundsscalartensorefts}, although no corresponding statement is currently available for the complete operator basis. 

\subsubsection{An example: EdGB gravity}\label{subsec:EdGB}
\begin{figure}[!h] 
     \centering
     \includegraphics[width = 0.5\textwidth]{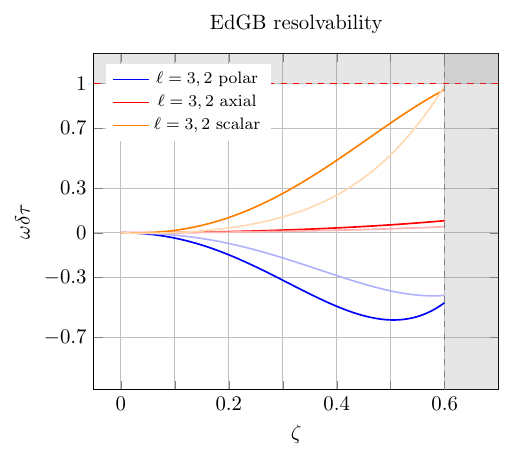}
    \caption{Resolvability at different values of the coupling in EDGB. The theory has three degrees of freedom. Graphs of the same color describe different values of $\ell$. There are cases when the decay-time shift is positive, but it is always smaller than $1$ within the trustworthy domain in $\zeta$.}
               \label{fig:EDGBris}
\end{figure}

Within scalar-tensor theories there is a clear example in which the Green's function poles migrate upwards in the complex plane:
\begin{equation}
    \mathcal{L}_{\text{EdGB}} = R - \frac{1}{2} \left(\partial \varphi \right)^2 + \frac{\alpha}{4} e^{\varphi} \mathcal{R}_{\text{GB}}^2 
\end{equation}
where we follow the notation in \cite{Salcedo1609}, set $G=1$ and do not normalize the fields canonically. Blázquez-Salcedo et al. \cite{Salcedo1609} find a spherically symmetric metric solution using two numerical methods, meaning they both:
\begin{enumerate}
    \item integrate fully numerically; 
    \item expand in $\zeta := 4\alpha/r_s^2$ small, up to $\mathcal{O}\left(\zeta^5\right)$ and then integrate. 
\end{enumerate}
For what concerns the series in $\zeta$, it could in principle happen that truncating the metric at $\mathcal{O}\left(\zeta^5\right)$ gives a very different result from the exact one at finite zeta. This depends on the radius of convergence of the series. Indeed, in \cite{Salcedo1609}, the results obtained with the second method deviate from the fully numerical ones for values of $\zeta$ around $\sim 0.3$. As a consequence, we trust the expansion: 
\begin{equation}
        \frac{\Im[\omega]}{\Im[\omega_{\rm GR}]} = 1 + I_1 \zeta + \cdots + I_5 \zeta^5 \qquad
        \frac{\Re[\omega]}{\Re[\omega_{\rm GR}]} = 1 + R_1 \zeta + \cdots + R_5 \zeta^5
\end{equation}
up to $\zeta \sim 0.3$ when using the values of $I_i$ and $R_i$ provided by \cite{Salcedo1609}. Computing $\omega \delta \tau$, the result for $n=0$, $m=0$ are the ones showed in Figure \ref{fig:EDGBris}.
While the shift can point both in the negative and positive direction in the complex plane, it remains unresolvable in all such cases. 

We now restrict attention to the axial gravitational sector, which decouples from the scalar perturbation on a static, spherically symmetric EdGB background. The polar gravitational and scalar perturbations instead form a coupled system. Determining their large-$\ell$ spectrum requires diagonalizing the leading eikonal potential and lies beyond the scope of this work. Our conclusions below therefore apply only to the axial QNM branch. We parametrize the EFT corrections using the functions $\Omega(\zeta)$ and $\lambda(\zeta)$ to make contact with the existing literature (for example \cite{Cardoso0912}):
\begin{equation}
    \omega \sim \Omega(\zeta) \left( \ell + \frac{1}{2} \right) - \frac{i}{2} |\lambda(\zeta)| + \mathcal{O}\left(\frac{1}{\ell}\right)
\end{equation}
Neither $\lambda$ nor $\Omega$ depend explicitly on $l$, therefore we can expand in $\zeta$ as: 
\begin{equation}
    \omega \sim \left(\Omega_{GR}  + \zeta \Omega_1 + \mathcal{O}(\zeta^2) \right) \left( \ell + \frac{1}{2} \right) - i \left(n+\frac{1}{2}\right) \left(\lambda_{\rm GR} + \zeta \lambda_1 + \zeta^2 \lambda_2 + \mathcal{O}(\zeta^3) \right) + \mathcal{O}\left( \frac{1}{\ell} \right)
\end{equation}
Assuming without loss of generality, $\zeta > 0$, in order to violate the bound, we need that the corrections $\lambda_i$ sum to a negative value. A violation of QNM stability reads, in the eikonal limit:
\begin{equation}
    -4\pi \left( \ell + \frac{1}{2} \right)\left(\frac{\zeta \lambda_1 + \zeta^2 \lambda_2 + \mathcal{O}(\zeta^3) }{\lambda_{\rm GR} + \zeta \lambda_1 + \zeta^2 \lambda_2 + \mathcal{O}(\zeta^3) } \right) + \mathcal{O}\left( \frac{1}{\ell} \right) > 1
\end{equation}
Now, assuming that $(\lambda_1, \dots, \lambda_{m-1})$ are zero, we have: 
\begin{equation}
     -  4\pi \left( \ell + \frac{1}{2} \right) \zeta^m  \left(\frac{\lambda_m + \zeta \lambda_{m+1} + \mathcal{O}(\zeta^{2}) }{\lambda_{\rm GR}} \right) + \mathcal{O}\left( \frac{1}{\ell} \right) > 1
\end{equation}
Therefore if $\ell$ tends to infinity at least as $\ell \sim 1/\zeta^{m+1}$ as $\zeta \rightarrow 0$, we get the condition to be verified: 
\begin{equation} \label{eq:ineq}
    -\frac{\lambda_m}{\lambda_{\rm GR}} \frac{4\pi}{\zeta}  > 1 + 4\pi \frac{\lambda_{m+1}}{\lambda_{\rm GR}}+ \mathcal{O}\left( \zeta \right)
\end{equation} 
Since the first term is dominant in the perturbative limit, we just need to verify that $\lambda_m < 0$. We do it by direct computation using the background metric (see Appendix \ref{app: C}). The result is:
\begin{equation}
        \lambda_1 = 0 \qquad \lambda_2 =-\frac{121907 }{8398080 \sqrt{3}} \frac{1}{r_s} < 0
\end{equation}
From which a violation of the bound is obtained whenever
\begin{equation}
    \left| \frac{12\sqrt{3} \pi\lambda_2}{\zeta}  \right| > \text{const.} +  \mathcal{O}(\zeta)
\end{equation}
Sending $\zeta$ to zero, this inequality can always be verified, but at the cost of letting $\ell$ grow outside the allowed regime of validity of the theory, since, in these units, the bound from perturbativity of the equations of motion is: 
\begin{equation}
    \ell \ll \frac{r_s^2}{\alpha} \sim \frac{1}{\zeta} \ll \frac{1}{\zeta^3}
\end{equation}
This fits in our previous analysis, since the only way to violate the bound would be to let $\ell$ grow well beyond the limits allowed by well-definiteness of the EFT.

\section{Comments on Einstein-Maxwell theories}\label{sec:EM}
Lastly, we comment on the case of Reissner-Nordstr{\"o}m black holes studied in \cite{Boyce2025}. Their Lagrangian, written making the energy scales explicit, is: 
\begin{equation}
    \begin{split}
          \mathcal{L}_{\text{EMEFT}} = & - \frac{1}{4} F_{\mu \nu}F^{\mu \nu} +   \frac{M_P^2}{2} R + \frac{M_P^2}{\Lambda^2}\left( \Tilde{c}_1 R^2 + \Tilde{c}_2 R^{\mu \nu} R_{\mu \nu} + \Tilde{c}_3 R^{\mu \nu \alpha \beta} R_{\mu \nu \alpha \beta} \right)      \\
          & + \frac{1}{\Lambda^2}\left(\Tilde{c}_4R F_{\mu \nu}F^{\mu \nu} +  \Tilde{c}_5 R^{\mu \nu} F_{\mu}^{\ \lambda} F_{\nu \lambda} + \Tilde{c}_6 R^{\mu \nu \alpha \beta}F_{\mu \nu} F_{\alpha \beta} \right) + \\
          &  + \frac{1}{\Lambda^2M_P^2} \left( \Tilde{c}_7 F_{\mu \nu}F^{\mu \nu} F_{\alpha \beta}F^{\alpha \beta} + \Tilde{c}_8 F_{\mu \nu}F^{\nu \lambda}F_{\lambda \alpha}F^{\alpha \mu} \right) 
    \end{split} 
\end{equation}
where $\Tilde{c}_i$ are the rescaled coefficients to be compared with $c_i$ in \cite{Boyce2025}. The background solution is chosen to be:
\begin{equation}
    ds^2 = -f(r)dt^2 + \frac{1}{g(r)} dr^2 + r^2 d\Omega^2 \qquad A = \psi(r) dt
\end{equation}
where the functions $f$, $g$ and $\psi$ are the ones describing Reissner-Nordstroem black holes plus EFT corrections that are obtained solving the background equations order by order: 
\begin{equation}
    \begin{split}
        & f(r) = 1-\frac{r_s}{r} + \frac{Q^2}{r^2} + \delta f \qquad g(r) = 1-\frac{r_s}{r} + \frac{Q^2}{r^2} + \delta g \\ 
        & \psi(r)  = -\frac{Q}{r} + \delta \psi
    \end{split}
\end{equation}
under the assumption that the mass $M = r_s/2G$ and charge $Q$ of the Black hole are not modified by the EFT operators, implying that the functions $\delta f$, $\delta g$ and $\delta \psi$ decay sufficiently fast for $r\rightarrow \infty$. The zeroth order surface gravity of the perturbed solution therefore is still:
\begin{equation}
    \kappa := \frac{\sqrt{ M^2 -Q^2}}{r_+^2} \qquad \text{where} \qquad r_+ := M + \sqrt{M^2 -Q^2}
\end{equation}
Considering dynamical perturbations over the background, in this case, the propagating degrees of freedom are four, encoded in the Zerilli-Moncrief master scalars $Z_{1,2}^{\pm}$. The master equations are written in \cite{Boyce2025} in a manner similar in spirit to the one presented above. 

The result for $\delta \tau$ in \cite{Boyce2025} is numerical and is computed for $l=2$ changing the value of surface gravity $\kappa$. Using their conventions ($G=1$ and $r_s=1$):
\begin{equation} \label{eq:RNbound}
    \Re[\omega_{\rm GR}] \delta \tau = \left( \rho_0(\kappa) d_0 + \rho_6(\kappa) d_6 + \rho_9(\kappa) d_9 \right)
\end{equation}
where $\rho_0$, $\rho_6$ and $\rho_9$ are numerically evaluated functions, while
\begin{equation}
\begin{split}
    & d_0 := \frac{1}{\Lambda^2}\left(\Tilde{c}_2 + 4\Tilde{c}_3 +  \Tilde{c}_5 +  \Tilde{c}_6 + 4\Tilde{c}_7 +  2\Tilde{c}_8 \right)\\ 
    & d_9 := \frac{1}{\Lambda^2}\left(\Tilde{c}_2 + 4\Tilde{c}_3 +   \Tilde{c}_5 +  2\Tilde{c}_6 +  \Tilde{c}_8 \right)\\
    & d_6 := \frac{\Tilde{c}_6}{\Lambda^2}
\end{split}
\end{equation}
$\rho_0$ and $\rho_9$ are strictly negative for the chosen values of $\kappa$, while $\rho_6$ reaches also positive values. In any case their order of magnitude is at most order $10$. 

Restoring natural units, the correct pre-factor of formula (4.1) in \cite{Boyce2025}, should be only related to $M$ and $\Lambda$ through the combination $1/(r_s\Lambda)^2$. This would suppress parametrically the numerical values of the decay-time shift, in accordance with our previous analysis.  

\section{Conclusion}\label{sec:Conclusion}
The stability bound proposed in \cite{Melville2401} asks that an EFT-corrected black hole ring down at least as fast as its GR counterpart or, if more slowly, by no more than about one oscillation period. We have shown that, under standard power counting, it does not yield new, model-independent constraints on the EFT Wilson coefficients. 

First, we have re-examined the motivations offered for the bound. The order-one tolerance in \eqref{eq:stabbound} is borrowed from the resolvability of time advances, but it cannot be applied to QNMs: the decay time is a parameter of a coherent classical waveform, and its uncertainty is set by the signal-to-noise ratio, parameter correlations, detector noise and waveform systematics, not by an occupation-independent floor of order $1/\Re[\omega]$. Without the tolerance, on the other hand, the heuristic that heavy physics pushes the poles deeper into the lower half-plane would demand $\delta\tau\le0$ for every mode, which is too strong: $\mathbf{Riem}^3$ shifts the decay times of the two parity sectors in opposite directions, so its coefficient would have to vanish. 

Second, even granting the tolerance, the bound adds no new information. The dimensionless shift $\Re[\omega_{\rm GR}]\,\delta\tau$ grows with the angular momentum of the mode, but the EFT around a black hole is predictive only up to some $\ell_{\max}$, beyond which an infinite tower of higher-derivative operators contributes at the same order. When curvature and derivatives are controlled by a single scale, the only power counting with a known ultraviolet completion, we find $\ell_{\max}\sim(r_s\Lambda)^2$, and the shift becomes of order one only there, where predictivity is lost. Our argument is parametric: with the large numerical coefficients of \cite{Melville2401}, the shift already becomes resolvable about an order of magnitude below $\ell_{\max}$, but there it is dominated by the two dimension-8 operators and the bound reduces to $c_1,c_2\gtrsim0$: a weaker form of the positivity that graviton scattering amplitudes already impose on the same coefficients \cite{Caron-Huot:2022ugt}. A parametrically resolvable shift requires instead a derivative scale $\Lambda_\nabla\gg\Lambda$. Even then, in pure gravity the bound again reduces to positivity, while in scalar-tensor theories it could in principle probe one family of dimension-10 operators, $(\R,\P,\J)=(2,2,2)$. This two-scale counting, however, has no known ultraviolet completion, and positivity restricts it severely: to $\Lambda_\nabla\lesssim\Lambda$ in GREFT, if the higher-spin gap sits at $\Lambda_\nabla$ \cite{Caron-Huot:2022ugt}, and to $\Lambda_\nabla\lesssim(\hat M_P/f)^{1/4}\Lambda$ in scalar-tensor theories \cite{Caron-Huot:2022ugt, Serra2205}. 

The explicit examples show this clearly. In EdGB gravity the shift is unresolvable at low $\ell$ in both parity sectors; in the eikonal analysis of the axial sector, even a formal double-scaling limit crosses the proposed threshold only at angular momenta beyond the validity of the EFT. In Einstein-Maxwell theory, the shifts computed in \cite{Boyce2025} carry a factor $1/(r_s\Lambda)^2$, just as the general scaling predicts. Along the way, we have extended the regime-of-validity analysis of \cite{de_Rham2007, Chen2112} to the full operator tower and to two-scale power counting (Section~\ref{sec:4} and Appendix~\ref{app: B}): it tells us up to which $\ell$ ringdown predictions are under control, and which operators dominate near that edge.

Several directions remain open. An explicit computation of the QNM shifts induced by the family $\nabla^{2\J}\mathbf{Riem}^{\J+4}$ would test our power counting beyond dimension eight. In Einstein--Maxwell theory, an analysis of the eikonal limit would extend the results of \cite{Boyce2025} to large $\ell$, and could be used to revisit the results in \cite{dirusso2025scalarquasinormalmodesreissnernordstrom}; we expect no qualitative difference from GREFT. More interesting is a general WKB treatment of coupled systems. The approach of \cite{Hui2311}, which we follow in Appendix~\ref{app:A}, requires the off-diagonal terms of the potential matrix to be subdominant. This condition fails when the leading-order potentials are nearly degenerate. 
Extending that approach would put the analyses of \cite{Glampedakis1906, Silva2020, Silva_2023} on a general footing, valid for arbitrary corrections to the action. 

More broadly, consistency conditions for EFTs on black-hole backgrounds remain largely uncharted. Indeed, microcausality and positivity without Lorentz invariance have only recently begun to be understood \cite{Hui2025, Creminelli:2025rxj}. Causality constraints from the propagation of waves on black-hole backgrounds do not yet stand on the same footing as flat-space positivity: in four dimensions the asymptotic causal structure of a black hole is insensitive to higher-derivative corrections \cite{Bucciotti:2026svg}. A notion of positivity adapted to black-hole backgrounds would be of interest in its own right, and would sharpen the priors on beyond-GR templates in ringdown tests of GR.

\section{Acknowledgment}
The authors thank Bruno Bucciotti, Antonio de Felice, Walter del Pozzo, Adrien Kuntz Scott Melville and Francesco Serra for fruitful discussions, and Francesco Serra for comments on the draft. The authors are partly supported by the Italian MUR under contract 20223ANFHR (PRIN2022).

\newpage

\appendix

\section{WKB for coupled systems of equations} \label{app:A}
The most general scenario for the equations of motion is provided by a system written as: 
\begin{equation}
    \frac{d^2 \Psi}{dr^{*2}} + \mathbf{V}(\Lambda, \omega, \ell; r_*)  \Psi = 0
\end{equation}
we follow the discussion in \cite{Hui2311} where the authors generalize the WKB quantization condition \eqref{eq:quantcond} to coupled systems\footnote{Other methods to tackle coupled systems of equations in the eikonal sector, though limited to specific types of corrections, can be found in \cite{Silva2020, Glampedakis1906}.}. Notice that it is not possible to diagonalize the potential without introducing first derivative terms. \\ 

\noindent The assumptions are: 
\begin{itemize}
    \item $V_{ij}(\Lambda, \omega, \ell; r_*\rightarrow \pm \infty) \rightarrow \Tilde{k}_{i \ \pm}(\Lambda, \omega, \ell) \delta_{ij}$  this is to reproduce the GR behavior very far away from the Black hole, and to ensure that the master scalars behave as a waves. 
    \item The functions in $\Psi(r_*)$ respect the QNM boundary conditions: 
    \begin{equation}
        \psi_i(r_* \rightarrow \pm \infty) \propto e^{\pm i \sqrt{ \Tilde{k}^i_{\pm} } r_*}  
    \end{equation}
    \item The potential goes sufficiently fast to the constant value:
    \begin{equation}
        \lim_{r_* \rightarrow \pm \infty} | r_* \mathbf{V}(\Lambda, \omega, \ell; r_*) - r_* \Tilde{k}_{i \ \pm}(\Lambda, \omega, \ell) \mathbb{I}| = 0 
    \end{equation}
    \item The off diagonal terms are perturbative: $\delta V_{ij}(\Lambda, \omega, \ell; r_*)$ for $i\neq j$ and their products $\delta V_{ij} \delta V_{jk}$ are sub-leading with respect to the perturbative corrections of the diagonal terms $\delta V_{ii}$. 
\end{itemize}
The last condition is necessary to make the procedure in \cite{Hui2311} consistent\footnote{The diagonalization of the zeroth order matrix in Eq. (4.2) in \cite{Hui2311}, does not imply that the first matrix element on the diagonal does not present an $x$ term. This is guaranteed perturbatively only if the off-diagonal terms are sub-dominant.}.  \\ 

There is a set of quasinormal modes for each eigenvalue of the potential matrix. Assuming that the off-diagonal terms in the potential matrix are perturbative implies that the diagonal terms in the potential matrix coincide with the eigenvalues up to second order correction in the perturbative expansion. For example: 
\begin{equation}
    \begin{pmatrix}
        V_{11}(r_*) & \delta V_{12}(r_*) \\ 
        \delta V_{21}(r_*) & V_{22}(r_*)
    \end{pmatrix} \overset{\text{diag.}}{\longrightarrow} \begin{pmatrix}
        V_{11}(r_*) + \frac{\delta V_{12} \delta V_{21}}{\Delta_{12}} & 0 \\ 
        0 & V_{22}(r_*) - \frac{\delta V_{12} \delta V_{21}}{\Delta_{12}}
    \end{pmatrix}\qquad \Delta_{12} \equiv V_{11} - V_{22}
\end{equation}
Therefore, under the assumptions that we stated above, we can use directly the quantization condition obtained in \cite{Hui2311}: 
\begin{equation} \label{eq:quantcond2}
   \frac{i V_{ii}(r_{\text{max}},\omega)}{\sqrt{2V_{ii}^{(2)}(r_{\text{max}},\omega)}} = n + \frac{1}{2} \qquad n \in \mathbb{N}_0 \qquad \forall i\in 1\dots N
\end{equation}
Since only two of the $N$ degrees of freedom come from the two helicity states of the GR graviton, we generically expand the potential as: 
\begin{equation}
    V_{ii}(r^{\text{EFT}}_{\text{max}}, \omega^{\text{EFT}}) = \omega^2_{i} - V_{\text{LO}, i}(\ell; r_*) + \delta V_i(\Lambda, \omega, \ell; r_*)
\end{equation}
where LO stands for \say{leading order} and can also come from the additional d.o.f. by itself. The term $\omega^2$ is always present for all the d.o.f., due to the presence of kinetic terms. Following the steps in the previous section, we obtain: 
\begin{equation}\label{eq:resolvabilitycoupledleading}
      \Delta V := \sqrt{2 V^{(2)}_{\text{LO}, i}}\left(  -\frac{\delta V_i}{\sqrt{2 V^{(2)}_{\text{LO}, i}}} + \frac{V^{(3)}_{\text{LO}, i}}{4\left(V^{(2)}_{\text{LO}, i}\right)^2} \delta V_i^{(1)} - \frac{\delta V_i^{(2)} }{4V^{(2)}_{\text{LO}, i}} \right).
\end{equation}
In order to use this WKB formula for coupled sets of equations, we should check if, and when, the requirements are satisfied. In particular, we need to guarantee that the product of mixing potential perturbation is sub-leading with respect to the perturbations of the diagonal ones, schematically meaning: 
\begin{equation} \label{eq:potentialhierarcy}
    \delta V_{i}\gg \frac{\delta V_{ij} \delta V_{j i}}{\Delta_{ij}}.
\end{equation}
It can be easily shown, by writing down the perturbed equations of motion, that within the regime of validity of the equations (see Appendix \ref{app: B}), this condition is always satisfied for the even sector of parity-preserving shift-symmetric scalar-tensor theories. 

\section{Validity scales around black holes}\label{app: B}
\subsection{GREFT}
Let us assume that the tree level EFT Lagrangian included just a single operator of the kind: 
\begin{equation}
    \hat{\mathcal{L}}_{\text{tree}} =  \left( \frac{\nabla}{\Lambda_{\nabla}} \right)^{2\J} \left(\frac{\textbf{Riem}}{\Lambda^2} \right)^{\R} \equiv \frac{\mathcal{O}_{\text{JR}}}{\Lambda_{\nabla}^{2\J} \Lambda^{2\R}} \qquad \text{for} \quad \R \geq 3
\end{equation}
where from now on we use the capital letter J to denote the power in the derivative expansion, and R to denote the power in the Riemann expansion. It is clear that the mass dimension of $\mathcal{O}_{\text{JR}}$ is $\Delta = 2(\J+\R)$. We notice that in order to write scalar operators the powers of nabla have to be even. Moreover, for $\R=2$ there are no non-redundant and non-topological operators, while for $\R=3$ the only operators of such kind involve no powers in J. While, for $\R \geq 4$, J can be generic. This can be checked explicitly, for example using the plethystic exponential method \cite{Ruhdorfer1908}. \\

\begin{figure}
    \centering
    \includegraphics[width=0.6\textwidth]{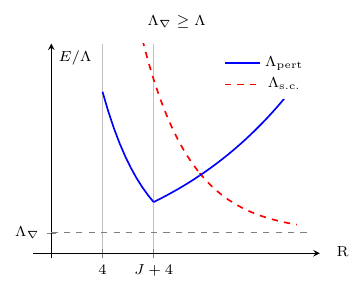}
             \caption{Relation between the strong coupling scale and the scale at which perturbativity breaks down for the allowed hierarchy between $\Lambda$ and $\Lambda_{\nabla}$.}
    \label{fig:relevant scales} 
\end{figure}

The graviton equations of motion will be generically modified by $\mathcal{O}_{\text{JR}}$ in the following way: 
\begin{equation} \label{eq:pertEOM}
    \Box h + \sum_n\sum_{i=0}^2 \Lambda^2 \left[\left( \frac{\partial}{\Lambda} \right)^{2-i}\left( \frac{\nabla}{\Lambda_{\nabla}} \right)^{2\J-n} \left( \frac{\overline{\textbf{Riem}}}{\Lambda^2} \right)^{\R-2}\right] \left(\frac{\partial}{\Lambda_{\nabla}}\right)^{n}\left( \frac{\partial}{\Lambda} \right)^{2+i} h = 0 
\end{equation}
where $n$ denotes the number of derivatives coming from the $\nabla$ expansion acting on $h$. To see clearly why this is true, it suffice to expand $\mathcal{O}_{\text{JR}}$ to second order in $h$: 
\begin{equation} \label{eq:expandedL}
   \mathcal{L}^{(2)}  \sim h \Box h  + \Lambda^2 \left( \frac{\nabla}{\Lambda_{\nabla}} \right)^{2\J} \left[ \left(\frac{\overline{\textbf{Riem}}}{\Lambda^2} \right)^{\R-2} \frac{1}{\Lambda^4} \partial^2 h \partial^2 h \right]
\end{equation}
where now $\nabla$ can act both on the background terms and on the perturbations. It was argued in \cite{Chen2112} that, due to the symmetries of the Riemann tensor, the number $n$ cannot be generic. In detail, notice that every $\partial$ and $\nabla$ carry a Lorentz index, while every $\overline{\textbf{Riem}}$ carries four Lorentz indices. If, in the EFT correction of \eqref{eq:pertEOM}, two $\partial$ were contracted, the corresponding contribution in the equation of motion would be zero, since to zeroth order
\begin{equation} \label{eq:onshell}
    \Box h \approx  0. 
\end{equation}
Given the on-shell condition \eqref{eq:onshell}, it is possible to contract only at most two $\partial$ with one $\overline{\textbf{Riem}}$, otherwise due to the anti-symmetry of its indices the contraction would yield zero: 
\begin{equation}
    \overline{\textbf{Riem}}_{\alpha \beta \mu \nu} \partial^{\alpha} \partial^{\beta} \partial^{\mu} = 0
\end{equation}
This means, that not only $n$ has to be smaller than $2\J$ by construction, but also: 
\begin{equation}
    n + 2 + i \leq 2-i + 2\J - n + 2(\R-2) \implies n \leq \R+\J- 2- i
\end{equation}
or
\begin{equation}
   0 \leq n \leq \min \left( 2\J ,  \R+\J-2-i\right) 
\end{equation}
An exceptional case is $(\R=3, \J=0)$, for which $i = 1$ and $n=0$ necessarily, since the derivative terms are at most of third order: 
\begin{equation}
    \textbf{Riem}^3 \supset h^{\mu \nu}\nabla_{\beta} R_{\alpha \gamma \lambda ( \nu} \partial^{\lambda}\partial^{(\gamma}\partial^{\beta)} h_{\mu)}^{\ \ \alpha}.
\end{equation}
Notice, moreover, that once the equations are reduced, time inversion and rotational invariance imply the absence of $\omega^3$ or $\ell^3$ terms in the eikonal limit, this explains the power of $\ell$ in \eqref{eq:eikonalshift}. We need to study two cases:
\begin{itemize}
    \item $\J > \R- 2 - i , \ \R\geq 4  \implies n \leq \R+\J-2 - i$, 
    \item $\J \leq \R-2 - i, \ \R\geq 4  \implies n \leq 2\J$.
\end{itemize}
A necessary condition for perturbativity of the equations of motion is that the EFT correction in \eqref{eq:pertEOM} is small compared to the GR (leading) contribution this will give an estimate for $\Lambda_{\text{pert}}$. Since we want to obtain the most stringent bound, we evaluate the background quantities using the smallest possible value of $r$, meaning $r_s$: $ \nabla \sim \frac{1}{r_s}$ and $\overline{\textbf{Riem}} \sim \frac{1}{r_s^2}$ and without loss of generality, we assume that all partial derivatives act as time derivative returning powers of $\omega$. Using \eqref{eq:pertEOM}, the necessary condition for the perturbativity of EOM therefore becomes: 
\begin{equation}
    \left(\frac{1}{r_s\Lambda} \right)^{2-i}\left(\frac{1}{r_s \Lambda_{\nabla}} \right)^{2\J-n} \left( \frac{1}{r_s^2 \Lambda^2} \right)^{\R-2} \left(\frac{\omega}{\Lambda_{\nabla}} \right)^{n} \left(\frac{\omega}{\Lambda} \right)^{i} \ll 1 \implies \omega \ll \frac{1}{r_s} \left[\left(r_s \Lambda_{\nabla} \right)^{2\J} \left( r_s\Lambda \right)^{2\R-2}\right]^{\frac{1}{n+i}}
\end{equation}
For $r_s$ fixed, the most stringent bound on $\omega$ is obtained for the largest even $n+i$ possible, meaning:
\begin{equation}
    \Lambda_{\text{pert}} = \begin{cases}
        \Lambda(r_s\Lambda)^3 & \J=0, \ \R = 3 \\
        \Lambda \left(r_s \Lambda_{\nabla}\right)^{\frac{2\J}{\R+\J-2}} \left( r_s\Lambda \right)^{\frac{\R-\J}{\R+\J-2}} & \J > \R- 4, \ \R\geq 4 \\ 
       \Lambda \left(r_s \Lambda_{\nabla} \right)^{\frac{\J}{\J+1}} \left( r_s\Lambda \right)^{\frac{\R-\J-2}{\J+1}} & \J \leq \R-4, \ \R\geq 4 
        \end{cases}
\end{equation}
As shown in Figure \ref{fig:relevant scales}, the minimum scale is obtained for $\R=\J+4$, from which the results in the main section. It is useful to compare this to what obtained in \cite{Chen2112}. Secondly, the smallest strong coupling scale can be evaluated from the $2 \longrightarrow (\R-2)$ scattering generated by: 
\begin{equation}
    \mathcal{L} \supset \Lambda^2 M_P^2 \frac{\partial^{2\J} (\partial^2 h)^{\R}}{ \Lambda_{\nabla}^{2\J} \Lambda^{2\R} M_P^{\R}}
\end{equation}
providing the bound: 
\begin{equation}
    \frac{E^{2(\J+\R)}}{\Lambda_{\nabla}^{2\J} \Lambda^{2\R}} \frac{\Lambda^2}{M_P^{\R-2}} \ll  (4\pi)^{\text{R-2}}E^{4-\R}.
\end{equation}
From this: 
\begin{equation}
   \Lambda_{\text{s.c.}}(\R, \J) = \Lambda \left( \frac{4 \pi M_P}{\Lambda} \right)^{\frac{\R-2}{2\J+3\R-4}} \left(\frac{\Lambda_{\nabla}}{\Lambda} \right)^{\frac{\J}{\J+3\R/2-2}}
\end{equation}
To make the relationship among scales clear, we depict their behavior in Figure \ref{fig:relevant scales}. The strong coupling scale for a generic EFT is provided by the inferior of $\Lambda_{\text{s.c.}}(\R, \J)$, which is obtained by operators in the limit $\J \rightarrow \infty$ (that are always present in a generic EFT). This is how to derive the results in Table \ref{tab:1}.  

\subsection{Scalar-tensor theories}
The logic is similar to the previous one. We include here the possibility of gravity being external to the dynamics of additional degrees of freedom at low energy \cite{Serra2205}. In practice we consider operators of the following type: 
\begin{equation} \label{eq: stlag}
   \frac{\Lambda^4}{\chi^2} \hat{\mathcal{L}}_{\text{tree}} = \frac{\Lambda^4}{\chi^2}\left( \frac{\nabla}{\Lambda_{\nabla}}\right)^{\J}\left( \frac{\textbf{Riem}}{\Lambda^2}\right)^{\R}\left(\frac{\chi \partial \phi}{\Lambda^2} \right)^{\P}\equiv \frac{\Lambda^4}{\chi^2}\frac{\mathcal{O}_{\text{JRP}}}{\Lambda_{\nabla}^{\J}\Lambda^{ 2\R+ 2\P}}
\end{equation}
where $\J+ \P$ must be even to ensure Lorentz invariance. The explicit structure of the equations of motion is similar to the one in GREFT, apart from the fact that now we are considering a coupled set of two equations, one for the graviton and one for the scalar field, in case of absence of parity breaking interactions. Notice that since mixing term are also present, in order to satisfy the perturbativity of the complete set, we also need to verify that the product of the two mixing is sufficiently small. Though, it is easy to check that satisfying the perturbativity of non-mixing terms is a necessary and sufficient condition. The two requirements are: 
\begin{equation}\label{eq:stcond}
    \begin{split}
        & \omega \ll \frac{1}{r_s} \left[ \left(\frac{\hat{M}_P^2}{f^2} \right)(r_s\Lambda_{\nabla})^{\J}(r_s\Lambda)^{2\P+ 2\R-2} \right]^{\frac{1}{n + i}} \qquad \quad \text{$h$ eq.} \\ 
        & \omega \ll \frac{1}{r_s} \left[(r_s\Lambda_{\nabla})^{\J} (r_s\Lambda)^{2\P+ 2\R-4} \right]^{\frac{1}{n+i-1}} \qquad \qquad \qquad \text{$\varphi$ eq.}
    \end{split}
\end{equation}
where $f:= \Lambda/\chi$, and where we used the fact that the dominant $\bar{\phi}$ contribution provides the most stringent bound. In both cases, the conditions $\Box h \approx 0$ and $\Box \varphi \approx 0$ imply conditions on the values of $n$ allowed. In particular: 
\begin{equation}
    \begin{split}
        & 0 \leq n \leq \min(\J,\R+ (\J + \P)/2 -2-i )  \qquad i=0, 1, 2 \qquad \text{$h$ eq.} \\ 
        & 0 \leq n \leq \min(\J, \R + (\J+ \P)/2 - 1- i) \qquad i = 0,1 \qquad \quad \text{$\varphi$ eq.} 
    \end{split}
\end{equation}
For the $h$ equation we need $\R\geq 2$ if $\P\geq 1$ and $\R\geq 4$ if $\P = 0$, while for the $\varphi$ equation, we must impose $\P \geq 2$. \\

Concerning the $h$ sector, for the same reasons as in the GREFT case, except for $\R=3$, $\J=0$ the minimum scale is obtained using $\R=\frac{\J-\P+8}{2}$ and $\P=0$. We obtain: 
\begin{equation}
    \Lambda_{\text{pert}}^{\text{min}
    } = \begin{cases}
        \Lambda(r_s \Lambda_{\nabla})  & \qquad  \Lambda_{\nabla} \lesssim  \Lambda \frac{\hat{M}_P}{f} (r_s\Lambda)  \\
        \Lambda \frac{\hat{M}_P}{f} (r_s\Lambda)^2 & \qquad \Lambda_{\nabla} \gg  \Lambda \frac{\hat{M}_P}{f} (r_s\Lambda).
    \end{cases}
\end{equation}
From the $\varphi$ sector $\Lambda_{\text{pert}}^{\text{min}} = \Lambda (r_s \Lambda_{\nabla})$, which exhausts all the possibilities. 

\section{Details on EdGB}\label{app: C}
We compute here $\lambda_m$ using the components of the metric. From \cite{Cardoso0912} we read off the formula for the Lyapunov exponents: 
\begin{equation}
    \lambda(\zeta) = \sqrt{ \frac{- r^2}{2f}  \sqrt{fg} \frac{d}{dr} \left( \sqrt{fg} \frac{d}{dr}\left( \frac{f}{r^2}\right) \right) }_{r=r_c}
\end{equation}
where $f(r, \zeta)$ and $g(r, \zeta)$ are the metric components in the modified Black hole solution: 
\begin{equation}
    ds^2 = -f(r, \zeta)dt^2 + \frac{1}{g(r, \zeta)} dr^2 + r^2 d \Omega^2
\end{equation}
and $r_c$ is new the location of the light-ring, obtained solving the equation: 
\begin{equation}
    2f(r_c)-r_c \left( \frac{d}{dr}f \right)_{r=r_c} = 0 
\end{equation}
Expanding both $f$ and $g$ in $\zeta$ it is possible to solve perturbatively Einstein's equations. The resulting $f_i$ and $g_i$ can be found in the additional material of \cite{Pierini2305} up to order six. Before expanding $\lambda$, we need to find the position of the light-ring, which can also be solved perturbatively. In particular, with $r(\zeta) = r_0 + \zeta r_1 + \zeta^2 r_2 + \mathcal{O}(\zeta^3)$ one obtains the following set of equations up to second order: 
\begin{equation}
\begin{split}
        2 f_0(r_0) -r_0 f'_0(r_0) = & 0 \qquad \text{0th-order} \\
        2 f_1(r_0) + r_1 f'_0(r_0) -r_0 f'_1(r_0) -r_0r_1 f''_0(r_0)  = & 0 \qquad \text{1st-order}\\
        r_2 f'_0  + 2f_2 + r_1f_1'-r_0\left(r_2 f''_0 + \frac{r_1^2}{2} f_0''' + f_2' + r_1f_1''  \right)   = & 0 \qquad \text{2nd-order} \\
       & \vdots
\end{split}
\end{equation}
where the prime denotes the radial derivative and all the functions are evaluated in $r_0$. Solving order by order one obtains: 
\begin{equation}
        r_0  = \frac{3}{2}r_s \qquad  r_1  = 0 \qquad  r_2  = \frac{4219}{207360} r_s
\end{equation}
and $\lambda(\zeta)$ to second order
\begin{equation}
    M \lambda(\zeta) = \frac{1}{3 \sqrt{3}} - \zeta \frac{ r_1 }{9 \sqrt{3}  M} -\zeta^2\frac{ \left( 38880 r_1^2/ M^2+29160 r_2 /M +2623\right)}{262440 \sqrt{3}}+ \mathcal{O}(\zeta^3)
\end{equation}
from which one can read the coefficient of $\zeta^2$:
\begin{equation}\label{eq:lambda}
 \lambda_2  = -\frac{121907}{8398080 \sqrt{3} } \frac{1}{r_s} < 0.
\end{equation}
The consequences of $\lambda_2$ being negative are outlined in Section \ref{subsec:EdGB}. This computation makes use of background quantities only, discarding the contribution of the leading effective operators on the effective metric. This is justified as far as EFT corrections to the effective metric can be eliminated by a diffeomorphism, which is the case at $\mathcal{O}(\zeta^2)$.

\bibliographystyle{jhep}
\bibliography{main.bib} 

\end{document}